\documentclass[11pt]{article}
\usepackage{amssymb}
\usepackage{amsmath}
\begin{document}
\title{{\bf{ Exact Differential and Corrected Area Law for Stationary Black Holes in Tunneling Method}}}
\author{
  {\bf {\normalsize Rabin Banerjee}$
 $\thanks{E-mail: rabin@bose.res.in}  and  {\normalsize Sujoy Kumar Modak}$ $\thanks{E-mail: sujoy@bose.res.in} }\\
 {\normalsize S.~N.~Bose National Centre for Basic Sciences,}
\\{\normalsize JD Block, Sector III, Salt Lake, Kolkata-700098, India}
\\[0.3cm]
}
\date{}

\maketitle
\begin{abstract}
We give a new and conceptually simple approach to obtain the ``first law of black hole thermodynamics'' from a basic thermodynamical property that entropy ($S$) for any stationary black hole is a state function implying that $dS$ must be an exact differential. Using this property we obtain some conditions which are analogous to Maxwell's relations in ordinary thermodynamics. From these conditions we are able to explicitly {\it calculate} the semiclassical Bekenstein-Hawking entropy, considering the most general metric represented by the Kerr-Newman spacetime. We extend our method to find the corrected entropy of stationary black holes in (3+1) dimensions. For that we first calculate the corrected Hawing temperature considering both scalar particle and fermion tunneling beyond the semiclassical approximation. Using this corrected Hawking temperature we compute the corrected entropy, based on properties of exact differentials. The connection of the coefficient of the leading (logarithmic) correction with the trace anomaly of the stress tensor is established . We explicitly calculate this coefficient for stationary black holes with various metrics, emphasising the role of Komar integrals.         
\end{abstract}



\section{{\bf Introduction:}}
Black holes are one of the most fascinating parts of theoretical, astrophysical and cosmological physics ever since Einstein's discovery of the theory of general relativity of gravitation. They are very important members of the universe. Because of their huge gravitational force no objects, not even light, can escape from them. There exists a region called `event horizon' beyond which all objects are strongly attracted towards the centre of a black hole leaving absolutely no chance for them to crossover the event horizon to the outer region. So they are completely isolated from the rest of the universe and have absolute zero temperature. However this is one part of black hole physics where everything is treated classically, but one has to check what happens when quantum effects are taken into account.

The inspiration of incorporating quantum theory for black holes is present within classical gravity itself. The four laws of ``{\it black hole mechanics}'' derived by Bardeen, Carter, Hawking \cite{Bardeen} are closely similar to the ``laws of thermodynamics'' if black holes are allowed to have some temperature. Around the same time of the above work, Bekenstein argued for black hole entropy based on simple aspects of thermodynamics \cite{Beken} which require that entropy of the universe cannot decrease due to the capture of any object by black holes. For making the total entropy of the universe at least unchanged, a  black hole should gain the same amount of entropy which is lost from the rest of the universe. Bekenstein then gave some heuristic arguments to show that black hole entropy must be proportional to its horizon area. He also fixed the proportionality constant as $\frac{ln2}{8\pi}$. The idea of Bekenstien was given a solid mathematical ground when Hawking incorporated quantum fields moving in a background of classical gravity and showed that black holes do emit particles having a black body spectrum with physical temperature $\frac{\hbar\kappa}{2\pi}$, where $\kappa$ is the surface gravity of a black hole \cite{Hawk}. Knowing this expression of black hole temperature (``{\it Hawking temperature}'') one can make an analogy with the `first law of black hole mechanics' and the `first law of thermodynamics' to identify entropy as $S=\frac{A}{4}$, where $A$ is the horizon area of the black hole. Thus it was proven that Bekenstein's constant of proportionality was incorrect and the new proportionality constant is $\frac{1}{4}$. The work of Bekenstein and Hawking thereby leads to the semiclassical result for black hole entropy encapsuled by the Bekenstein-Hawking area law, given by 
\begin{eqnarray}
S_{\textrm {BH}}=\frac{A}{4}.
\label{1.1}
\end{eqnarray}

Thereafter a lot of effort has been made for studying thermodynamic aspects of black holes. Indeed there are several approaches to calculate Hawking temperature and entropy of a black hole. Among these a simple and physically intuitive picture is provided by the tunneling mechanism \cite{Wilczek, Cai, Paddy, Kern}. It has two variants namely null geodesic method \cite{Wilczek, Cai} and Hamilton-Jacobi method \cite{Paddy, Kern}. Recently in \cite{Majhiflux}, Hawking flux from the tunneling mechanism has been derived which shows that black holes have perfect black body spectrum with the correct Hawking temperature. In tunneling method pair creation occurs just inside the event horizon where one mode moves towards the centre of the black hole while the other mode just tunnels through the event horizon to the outer region and reaches infinity.

Besides temperature, there have been various studies related to the obtention of entropy.  Although till now there is no microscopic description of black hole entropy, several approaches have shown that the semiclassical Bekenstein-Hawking entropy (\ref{1.1}) undergoes corrections. These approaches are mainly based on field theory \cite{Fursaev}, quantum geometry \cite{Partha}, statistical mechanics \cite{Das}, Cardy formula \cite{Carlip}, brick wall method \cite{Hooft} and tunneling method \cite{R. Banerjee, Majhi1, Modak, Majhitrace}. But none of these is successful to include all the black hole spacetimes. 

In this paper we construct a different framework for studying entropy using a basic property of ordinary thermodynamics that ensures entropy ($S$) must be a {\it state function}. This naturally yields the `first law of black hole thermodynamics' where one does not need the first law of black hole mechanics. The fact that $dS$ is an exact differential gives three integrability conditions which are analogous to {\it Maxwell's equations} in ordinary thermodynamics.  Unlike the usual approach where the Bekenstein-Hawking entropy is read-off by a comparison of the first law of thermodynamics with the first law of black hole mechanics, we are able to directly calculate the entropy by taking all work terms into consideration. It is revealed that although the work terms have some role to play in between, they do not contribute to the final result for the semiclassical Bekenstein-Hawking entropy. Our analysis is performed for a general black hole defined by the Kerr-Newman metric. The main strength of our approach, however, lies in finding the corrections to the semiclassical value of entropy. For this we first calculate corrections to semiclassical Hawking temperature using both scalar particle and fermion tunneling in the Kerr-Newman spacetime. Equivalent results are obtained. We also find that the corrected Hawking temperature, as calculated by tunneling mechanism, has several arbitrary coefficients. We determine these coefficients by demanding that the corrected entropy ($S_{\textrm {bh}}$) of a stationary black hole also has to be a {\it state function}. The integrability conditions (analog to Maxwell's relations) on $dS_{\textrm {bh}}$ fix most of the coefficients. Then following the usual technique to solve an exact differential equation, we calculate the corrected entropy for the Kerr-Newman black hole. In the limiting case the whole analysis is valid to give the corrected entropy for other black holes, as for example (i) Kerr, (ii) Reissner-Nordstrom and (iii) Schwarzschild black hole. The general form of the corrected entropy includes logarithmic terms and inverse area terms as leading and next to leading order corrections. However in the expression of the corrected entropy, there is one arbitrary coefficient present with each correctional term.

The remainder of this paper then deals with fixing the coefficient ($\tilde\beta_1$) of the logarithmic term. We successfully fix this coefficient for all spacetimes. It is related to the trace anomaly of the stress tensor. The concept of Komar conserved quantity corresponding to a Killing vector plays a crucial role for the explicit calculation of $\tilde\beta_1$. We consider various stationary black hole spacetimes in (3+1) dimensions and perform an integration over the trace anomaly to give the final result for $\tilde\beta_1$. From our analysis it is revealed that $\tilde\beta_1$ is a pure number for both Schwarzschild and Kerr spacetime and, more importantly, the values are exactly equal. This is consistent since there is no difference in the dynamics for these two black holes as they only differ in their geometrical behaviour. For the other two charged black holes (Reissner-Nordstrom and Kerr-Newman) $\tilde\beta_1$ is not a pure number but in the limit $Q=0$ they reproduce the result for Schwarzschild and Kerr black holes respectively.

The paper is organised as follows. In Section 2 we deduce the `first law of black hole thermodynamics' from a different viewpoint by considering entropy as a state function and calculate the semiclassical Bekenstein-Hawking entropy for stationary black holes. In Section 3 both scalar particle and fermion tunneling is used to calculate the corrected Hawking temperature. Section 4 is devoted to find the general form of a corrected area law which is valid for all stationary spacetimes in (3+1) dimensions. In Section 5 the coefficient of the leading (logarithmic) correction to the area law is fixed. Section 6 is left for our conclusions and discussions. We give our notations and definitions in an appendix which also includes a very brief review of Komar conserved quantities.

\section{{\bf Exact differential and semiclassical Area Law }}

Long time back (1973) within the realm of classical general relativity Bardeen, Carter and Hawking gave the ``first law of black hole mechanics'' which states that for two nearby black hole solutions the difference in mass ($M$), area ($A$) and angular momentum ($J$) must be related by \cite{Bardeen}
\begin{eqnarray}
\delta M= \frac{1}{8\pi}{\kappa\delta A}+ \Omega_{\textrm H} \delta J.
\label{bhmech}
\end{eqnarray}
In addition some more terms can appear on the right hand side due to the presence of other matter fields. They found this analogous to the ``first law of thermodynamics'', which states, the difference in energy ($E$), entropy ($S$) and other state parameters of two nearby thermal equilibrium states of a system is given by
\begin{eqnarray}
dE= T dS+ {\textrm {``work terms''}}.
\label{thermo}
\end{eqnarray}
Therefore even in classical general relativity the result (\ref{bhmech}) is appealing due to the fact that both $E$ and $M$ represent the same physical quantity, namely total energy of the system. Although at that time this result was quite surprising as classically, temperature of black holes was absolute zero. So the identification of temperature with surface gravity, as shown by (\ref{bhmech}) and (\ref{thermo}), was meaningless. Consequently, identification of entropy with horizon area was inconsistent.

 However the picture was changed dramatically when Hawking (1975), incorporating quantum effects, discovered \cite{Hawk} that black holes do radiate all kinds of particles with a perfect black body spectrum with temperature $T_{\textrm H}=\frac{\kappa}{2\pi}$. From this mathematical identification of the Hawking temperature ($T_{\textrm H}$) with the surface gravity ($\kappa$) in (\ref{bhmech}), one is left with some analogy between entropy ($S$) and the area of the event horizon($A$), suggested by (\ref{bhmech}) and (\ref{thermo}). The result $S=\frac{A}{4}$ follows from this analogy.

For such an identification, the horizon area of a black hole is playing the ``mathematical role'' of entropy and does not have a solid physical ground. Also, this naive identification remains completely silent about the role of ``work terms''. But if one does not use this mathematical analogy, rather tries to {\it calculate} entropy, it may appear that these work terms might have some role to play. Therefore the role of these work terms is not transparent in the process of identifying entropy. Moreover, in this analysis one can obtain the ``first law of black hole thermodynamics'' only by deriving the ``first law of black hole mechanics'' and then identifying this with the ordinary ``first law of thermodynamics''.

Now we want to obtain the ``first law of black hole thermodynamics''  by directly starting from the {\it thermodynamical} viewpoint where one does not require the ``first law of black hole mechanics''. From such a law the entropy will be explicitly calculated and not identified, as usually done, by an analogy between (\ref{bhmech}) and (\ref{thermo}). For this derivation we interpret Hawking's result of black hole radiation \cite{Hawk} as 
\begin{itemize}
\item {{\it black holes are thermodynamical objects having mass ($M$) as total energy ($E$) and they are immersed in a thermal bath in equilibrium with physical temperature ($T_{\textrm H}$)}.} 
\end{itemize}
Therefore following the ordinary ``first law of thermodynamics'' we are allowed to write the ``first law of black hole thermodynamics'' as 
\begin{eqnarray}
dM= T_{\textrm H}dS + {\textrm {``work terms on black hole''}},
\label{newlaw}
\end{eqnarray}
where $M$ is the mass of the black hole and $T_{\textrm H}$ is the Hawking temperature. Usually, without deriving the ``first law of black hole mechanics'' one is not able to find ``work terms on black hole'' exactly. But we can always make a dimensional analysis to construct these two terms as proportional to $\Omega_{\textrm H} dJ$ and $\Phi_{\textrm H} dQ$ where $J$ and $Q$ are the angular momentum and charge of the black hole. This is possible since the form of ``angular velocity ($\Omega_{\textrm H}$)'' and ``potential ($\Phi_{\textrm H}$)'' at the event horizon are known individually from classical gravity. These terms can be brought on the right hand side of (\ref{newlaw}) with some prefactors given by dimensionless constants `$a$' and `$b$', such that (\ref{newlaw}) becomes 
\begin{eqnarray}
dM=T_{\textrm H} dS+a\Omega_{\textrm H} dJ+ b\Phi_{\textrm H} dQ.
\label{4.11}
\end{eqnarray}
To fix the arbitrary constants `$a$' and `$b$' let us first rewrite (\ref{4.11}) in the form 
\begin{eqnarray}
dS=\frac{dM}{T_{\textrm H}}+(-\frac{a\Omega_{\textrm H}}{T_{\textrm H}})dJ+(-\frac{b\Phi_{\textrm H}}{T_{\textrm H}})dQ
\label{4.41}
\end{eqnarray}
From the principle of ordinary first law of thermodynamics one must interpret entropy as a {\it state function}. For the evolution of a system from one equilibrium state to another equilibrium state, entropy does not depend on the details of the evolution process, but only on the two extreme points representing the equilibrium states. This universal property of entropy must be satisfied for black holes as well. In fact the entropy of any stationary black hole should not depend on the precise knowledge of its collapse geometry but only on the final equilibrium state. Hence we can conclude that entropy for a stationary black hole is a state function and consequently $dS$ has to be an {\it exact differential}. As a result the coefficients of the right hand side of (\ref{4.41}) must satisfy the three integrability conditions  
\begin{align}
\frac{\partial}{\partial J}(\frac{1}{T_{\textrm H}})\big|_{M,Q}& =\frac{\partial}{\partial M}(-\frac{a\Omega_{\textrm H}}{T_{\textrm H}})\big|_{J,Q}\notag\\
\frac{\partial}{\partial Q}(-\frac{a\Omega_{\textrm H}}{T_{\textrm H}})\big|_{M,J}& =\frac{\partial}{\partial J}(-\frac{b\Phi_{\textrm H}}{T_{\textrm H}})\big|_{M,Q}\notag\\
\frac{\partial}{\partial M}(-\frac{b\Phi_{\textrm H}}{T_{\textrm H}})\big|_{J,Q}& =\frac{\partial}{\partial Q}(\frac{1}{T_{\textrm H}})\big|_{J,M}.
\label{4.71}
\end{align}
As one can see, these relations are playing a role similar to {\it Maxwell's relations} of ordinary thermodynamics. Like Maxwell's relations these three equations do not refer to a process but provide relationships between certain physical quantities that must hold at equilibrium.

The only known stationary solution of Einstein-Maxwell equation with all three parameters, namely Mass ($M$), Charge($Q$) and Angular momentum ($J$) is given by the Kerr-Newman spacetime. All the necessary information for that metric is provided in Appendix 7.1 and one can readily check that the first, second and third conditions are satisfied only for $a=1,~~a= b~$ and $~b=1$ respectively, leading to the unique solution $a=b=1$. As a result, (\ref{4.11}) immediately reduces to the standard form 
\begin{eqnarray}
dM=T_{\textrm H} dS+\Omega_{\textrm H} dJ+\Phi_{\textrm H} dQ,
\label{sthermo}
\end{eqnarray}
This completes the obtention of the ``first law of black hole thermodynamics'',  for a rotating and charged black hole,  without using the ``first law of black hole mechanics''.

One can make an analogy of (\ref{sthermo}) with the standard first law of thermodynamics given by $dE= TdS- pdV +\mu dN$. Knowing $E=M$ (since both represent the same quantity which is the energy of the system) one can infer the correspondence $-\Omega_{\textrm H}\rightarrow p,~J\rightarrow V,~\Phi_{\textrm H}\rightarrow \mu,~Q\rightarrow N$ between the above two cases. Indeed $\Omega_{\textrm H} dJ$ is the work done on the black hole due to rotation and is the exact analogue of the $-pdV$ term. Likewise the electrostatic potential $\Phi_{\textrm {H}}$ plays the role of the chemical potential $\mu$.   


It is now feasible to calculate the entropy by using properties of exact differentials. The first step is to rewrite (\ref{sthermo}) as
\begin{eqnarray}
dS=\frac{dM}{T_{\textrm H}}+(\frac{-\Omega_{\textrm H}}{T_{\textrm H}})dJ+(\frac{-\Phi_{\textrm H}}{T_{\textrm H}})dQ,
\label{sthermo1}
\end{eqnarray}
where $dS$ is now an exact differential.

 Any first order partial differential equation   
\begin{eqnarray}
df(x,y,z)=U(x,y,z)dx+V(x,y,z)dy+W(x,y,z)dz
\label{4.8}
\end{eqnarray}
is exact if it fulfills these integrability conditions
\begin{eqnarray}
\frac{\partial U}{\partial y}\big|_{x,z}=\frac{\partial V}{\partial x}\big|_{y,z};~~~\frac{\partial V}{\partial z}\big|_{x,y}=\frac{\partial W}{\partial y}\big|_{x,z};~~~\frac{\partial W}{\partial x}\big|_{y,z}=\frac{\partial U}{\partial z}\big|_{x,y}.
\label{4.9}
\end{eqnarray}
If these three conditions hold then the solution of (\ref{4.8}) is given by
\begin{eqnarray}
f(x,y,z)=\int{ Udx} +\int{Xdy}+\int{Ydz},
\label{4.12}
\end{eqnarray}
where
\begin{eqnarray}
X=V-\frac{\partial}{\partial y}{\int{Udx}}
\label{4.13}
\end{eqnarray}
and
\begin{eqnarray}
Y=W-\frac{\partial}{\partial z}[\int{Udx}+{\int Xdy}].
\label{4.14}
\end{eqnarray}
 Now comparing (\ref{sthermo1}) and (\ref{4.8}) we find the following dictionary
\begin{align}
(f\rightarrow S,~~x\rightarrow M,~~y\rightarrow J,~~z\rightarrow Q)\nonumber\\
(U\rightarrow\frac{1}{T_{\textrm H}},~~V\rightarrow\frac{-\Omega_{\textrm H}}{T_{\textrm H}},~~W\rightarrow\frac{-\Phi_{\textrm H}}{T_{\textrm H}}).
\label{4.15}
\end{align}
Using this dictionary and (\ref{4.12}), (\ref{4.13}) and (\ref{4.14}) one finds,
\begin{eqnarray}
S=\int{\frac{dM}{T_{\textrm H}}}+\int{XdJ}+\int{YdQ},
\label{4.19}
\end{eqnarray}
where
\begin{eqnarray}
X=(-\frac{\Omega_{\textrm H}}{T_{\textrm H}})-\frac{\partial}{\partial J}{\int\frac{dM}{T_{\textrm H}}}
\label{4.20}
\end{eqnarray}
and
\begin{eqnarray}
Y=(-\frac{\Phi_{\textrm H}}{T_{\textrm H}})-\frac{\partial}{\partial Q}[{\int\frac{dM}{T_{\textrm H}}}+{\int XdJ}].
\label{4.21}
\end{eqnarray}
In order to calculate the semiclassical entropy we need to solve (\ref{4.19}), (\ref{4.20}) and (\ref{4.21}). Note that all the ``work terms'' are appearing in the general expression of the semiclassical entropy of a black hole (\ref{4.19}). Let us first perform the mass integral to get
\begin{eqnarray}
\int\frac{dM}{T_{\textrm H}}=\frac{\pi}{\hbar}\left(2M{[M+(M^2-\frac{J^2}{M^2}-Q^2)^{1/2}}]-Q^2\right),
\label{4.22}
\end{eqnarray}
 where the expression (\ref{4.5}) has been substituted for $T_{\textrm H}^{-1}$. With this result one can check the following equality 
\begin{eqnarray}
\frac{\partial}{\partial J}\int\frac{dM}{T_{\textrm H}}=-\frac{\Omega_{\textrm H}}{T_{\textrm H}}
\label{4.23}
\end{eqnarray}
holds, where $\Omega_{\textrm H}$ is defined in (\ref{angv}). Putting this in (\ref{4.20}) it follows that $X=0$. Using (\ref{4.22}) one can next calculate, 
\begin{eqnarray}
\frac{\partial}{\partial Q}\int\frac{dM}{T_{\textrm H}}=-\frac{\Phi_{\textrm H}}{T_{\textrm H}},
\label{4.24}
\end{eqnarray}
 where $-\frac{\Phi_{\textrm H}}{T_{\textrm H}}$ is given in (\ref{4.7}). With this equality and the fact that $X=0$, we find, using (\ref{4.21}), $Y=0$. Exploiting all of the above results, the semiclassical entropy for Kerr-Newman black hole is found to be, 
\begin{eqnarray}
S=\int{\frac{dM}{T_{\textrm H}}}=\frac{\pi}{\hbar}\left(2M{[M+(M^2-\frac{J^2}{M^2}-Q^2)^{1/2}}]-Q^2\right)=\frac{A}{4\hbar}=S_{{\textrm {BH}}},
\label{4.25}
\end{eqnarray}
 which is the standard semiclassical Bekenstein-Hawking area law for Kerr-Newman black hole. The expression for the area ($A$) of the event horizon follows from (\ref{area}). Now it is trivial, as one can check, that all other stationary spacetime solutions, for example Kerr or Reissner-Nordstrom, also fit into the general framework to give the semiclassical Bekenstein-Hawking area law. Thus the universality of the approach is justified.

\section{{\bf Correction to semiclassical Hawking temperature}}

For convenience of our analysis let us first rewrite the original Kerr-Newman metric (given in Appendix 7.1) in the following form,
\begin{align}
ds^{2}  &  =-F(r,\theta)dt^{2}+\frac{dr^{2}}{\tilde g(r,\theta)}+K(r,\theta
)(d\phi-\frac{H(r,\theta)}{K(r,\theta)}dt)^{2}+\Sigma(r)d\theta^{2},%
\label{2.3}\\
F(r,\theta)  &  =\tilde f(r,\theta)+\frac{H^{2}(r,\theta)}{K(r,\theta)}=\frac{\Delta(r)\Sigma(r,\theta)}{(r^{2}+a^{2})^{2}-\Delta(r)a^{2}\sin^{2}\theta}
\nonumber
\end{align}
In course of finding the correction to the semiclassical Hawking temperature we follow the method developed in \cite{Majhi1}. Therefore the first aim is to isolate the `$r-t$' sector of the metric (\ref{2.3}) from the angular part. In a previous analysis for rotating BTZ black hole we did a similar work \cite{Modak}. The idea is to take the near horizon form of the metric and thereby redefine the angular part in such a way that the $r-t$ sector becomes isolated. This redefinition only changes the total energy of the tunneling particle \cite{Modak,Kerner1,Ang1} and does not affect the thermodynamical entities. In the case of Kerr-Newman black hole this issue is little more subtle since the metric coefficients also depend on $\theta$. However, because of the presence of an ergosphere, $\tilde f(r,\theta)$ in (\ref{2.1}) is positive on the horizon for two specific values of $\theta$, say, $\theta_0=0$ or $\pi$. For these two values of $\theta$ the ergosphere and the event horizon coincide. For the tunneling of any particle through the horizon of the Kerr-Newman black hole only these two specific values of $\theta$ are allowed. When we take the near horizon limit of the metric (\ref{2.3}) the value of $\theta$ is first fixed to $\theta_0$. The form of the metric near the horizon for fixed $\theta=\theta_{0}$\ is given by \cite{Kerner1},       
\begin{equation}
ds^{2}=-F'(r_{+},\theta_{0})(r-r_{+})dt^{2}+\frac{dr^{2}}{\tilde g'(r_{+},\theta_{0})(r-r_{+})}+K(r_{+},\theta_{0})(d\phi-\frac{H(r_{+},\theta_{0})}{K(r_{+},\theta_{0})}dt)^{2} 
\label{2.4}
\end{equation}
where,
\begin{eqnarray}
\frac{H(r_+,\theta)}{K(r_+,\theta)}=\frac{a}{r_{+}^{2}+a^{2}}=\Omega_{\textrm H}
\label{2.5}
\end{eqnarray}
is the angular velocity of the event horizon. A coordinate transformation
\begin{eqnarray} 
d\chi=d\phi-\Omega_{\textrm H}dt\implies\chi=\phi-\Omega_{\textrm H} t
\label{2.51}
\end{eqnarray}
 will take the metric (\ref{2.4}) into the desired form, %
\begin{equation}
ds^{2}=-F'(r_{+},\theta_{0})(r-r_{+})dt^{2}+\frac{dr^{2}}{\tilde g'(r_{+},\theta_{0})(r-r_{+})}+K(r_{+},\theta_{0})d\chi^{2}, 
\label{2.6}
\end{equation}
 where the `$r-t$' sector is isolated from the angular part $d\chi^2$.
Note that the `$r-t$' sector of the metric (\ref{2.6}) has the form,
\begin{eqnarray}
ds^2=-f(r)dt^2+{\frac{1}{g(r)}}{dr^2}, 
\label{2.7}
\end{eqnarray}
where
\begin{eqnarray}
f(r)=F'(r_{+},\theta_{0})(r-r_{+})
\label{2.71}\\
g(r)=\tilde g'(r_{+},\theta_{0})(r-r_{+}). 
\nonumber
\end{eqnarray}

\subsection{{\bf Scalar Particle tunneling}}

The massless particle in spacetime (\ref{2.6}) is governed by the Klein-Gordon equation
\begin{equation}
-\frac{\hbar^2}{\sqrt{-g}}{\partial_\mu[g^{\mu\nu}\sqrt{-g}\partial_{\nu}]\Phi}=0.
\label{2.8}
\end{equation}  
 In the tunneling approach we are concerned about the radial trajectory, so that only the $r-t$ sector (\ref{2.7}) of the metric (\ref{2.6}) is relevant. Note that in the analysis given in \cite{Majhi1}, for a Shwarzschild black hole, the structure of the `$r-t$' sector was similar to (\ref{2.7}). But it should be remembered that now we are dealing with a black hole having three parameters ({$M, Q, J$}). As a consequence a major difference will appear later on.

Equation (\ref{2.8}), with the background metric (\ref{2.7}) cannot be solved exactly. Therefore we start with the following standard WKB ansatz for $\Phi$ as
     
\begin{eqnarray}
\Phi(r,t)=exp[\frac{i}{\hbar}{{\cal S}(r.t)}], 
\label{2.9}
\end{eqnarray}
and substitute it in (\ref{2.8}) to yield,
\begin{eqnarray}
&&\frac{i}{\sqrt{f(r)g(r)}}\Big(\frac{\partial S}{\partial t}\Big)^2 - i\sqrt{f(r)g(r)}\Big(\frac{\partial S}{\partial r}\Big)^2 - \frac{\hbar}{\sqrt{f(r)g(r)}}\frac{\partial^2 S}{\partial t^2} + \hbar \sqrt{f(r)g(r)}\frac{\partial^2 S}{\partial r^2}
\nonumber
\\
&&+ \frac{\hbar}{2}\Big(\frac{\partial f(r)}{\partial r}\sqrt{\frac{g(r)}{f(r)}}+\frac{\partial g(r)}{\partial r}\sqrt{\frac{f(r)}{g(r)}}\Big)\frac{\partial S}{\partial r}=0.
\label{2.9a}
\end{eqnarray}
Then expanding the action ${\cal S}(r,t)$ in the powers of $\hbar$ 
\begin{eqnarray}
{\cal S}(r,t)={\cal S}{_0}(r,t)+  \sum_i{\hbar^i {\cal S}_i(r,t)},
\label{2.10}
\end{eqnarray}
and putting this in (\ref{2.9a}) one gets a set of differential equations for different order of $\hbar$ and those can be simplified to obtain,
\begin{eqnarray}
\hbar^0~:~\frac{\partial S_0}{\partial t}=\pm \sqrt{f(r)g(r)}\frac{\partial S_0}{\partial r},
\label{2.11}
\end{eqnarray}
\begin{eqnarray}
\hbar^1~:~&&\frac{\partial S_1}{\partial t}=\pm \sqrt{f(r)g(r)}\frac{\partial S_1}{\partial r},
\nonumber
\\
\hbar^2~:~&&\frac{\partial S_2}{\partial t}=\pm \sqrt{f(r)g(r)}\frac{\partial S_2}{\partial r},
\nonumber
\\
.
\nonumber
\\
.
\nonumber
\\
.
\nonumber
\end{eqnarray} 
and so on. Note that the $n$-th order solution is expressed by,
\begin{eqnarray}
\frac{\partial S_n}{\partial t}=\pm \sqrt{f(r)g(r)}\frac{\partial S_n}{\partial r},
\label{2.11a}
\end{eqnarray}
where ($n=~0,~i;~i= 1,2,...) $.

The most general form of semiclassical action in the original Kerr-Newman spacetime is given by
\begin{eqnarray}
{\cal S}_0(r,t,\theta,\phi)=-Et+ P_{\phi}\phi+ \tilde {\cal S}_0(r,\theta),
\label{2.12}
\end{eqnarray}
 where $E$ and $P_{\phi}$ are the Komar conserved quantities \cite{Komar} ({\it see} Appendix 7.2) corresponding to the two Killing vectors $\partial_t$ and $\partial_{\phi}$.
 In the near horizon approximation for fixed $\theta=\theta_0$ and using (\ref{2.51}) one can isolate the semiclassical action for the `$r-t$' sector as,
\begin{eqnarray}
{\cal S}_0(r,t)=-\omega t+ \tilde {\cal S}_0(r),
\label{2.13}
\end{eqnarray}
 where
\begin{eqnarray}
\omega=(E-P_{\phi}\Omega_{H})
\label{2.13a}
\end{eqnarray}
 is identified as the total energy of the tunneling particle. The solution for other ${\cal S}_i(r,t)$' s, subjected to a choice similar to (\ref{2.13}), can at best differ by a proportionality factor, since they satisfy generically identical equations as (\ref{2.11a}). The most general form of action including the contribution from all orders of $\hbar$ is then given by \cite{Majhi1, Modak, Majhitrace}
\begin{eqnarray}
{\cal S}(r,t)=(1+\sum\gamma_i \hbar^i) {\cal S}_0(r,t),
\label{2.14}
\end{eqnarray}
It is clear that the dimension of $\gamma_i$ is equal to the dimension of $\hbar^{-i}$. Let us now perform the following dimensional analysis to express these $\gamma_i$' s in terms of dimensionless constants. In (3+1) dimensions in the unit of $G= c= \kappa_B= \frac{1}{4\pi\epsilon_0}= 1$, $\sqrt\hbar$ is proportional to Plank length ($l_p$), Plank mass ($m_p$) and Plank charge ($q_p$) {\footnote {$l_p=\sqrt{{\frac{\hbar G}{c^3}}} , m_p=\sqrt{\frac{\hbar c}{G}} , q_p= \sqrt{c\hbar 4\pi\epsilon_0 }.$ }}. Therefore the most general term which has the dimension of $\hbar$ can be expressed in terms of black hole parameters as 
\begin{equation}
H_{\textrm {KN}}(M,J,Q)=  {a_1r^2_{+}}+a_2 {Mr_+}+a_3{M^2}+a_4{r_+ Q}+a_5{MQ}+a_6{Q^2}. 
\label{2.15}
\end{equation} 
Using this the action in (\ref{2.14}) now takes the form
\begin{eqnarray}
{\cal S}(r,t)=(1+\sum\frac{\beta_i\hbar^i}{H_{\textrm {KN}}^{i}}) {\cal S}_0(r,t).
\label{2.16}
\end{eqnarray}
  where $\beta_i$'s are dimensionless constants.

To find the solution for $S_0(r,t)$, let us put (\ref{2.13}) in the first partial differential equation in (\ref{2.11}) and integrate to obtain
\begin{eqnarray}
\tilde{{\cal S}_0}(r) =  \pm \omega\int_C\frac{dr}{\sqrt{f(r)g(r)}}
\label{2.17}
\end{eqnarray}
The + (-) sign indicates that the particle is outgoing (ingoing). Using the expression for ${\cal S}_0(r,t)$ from (\ref{2.13}) and (\ref{2.17}) one can write (\ref{2.16}) as
\begin{eqnarray}
{\cal S}(r,t)=(1+\sum\frac{\beta_i\hbar^i}{H_{\textrm {KN}}^{i}})(-\omega t \pm \omega\int_C\frac{dr}{\sqrt{f(r)g(r)}}) .
\label{2.18}
\end{eqnarray}
The solution for the ingoing and outgoing particle of the Klein-Gordon equation under the background metric (\ref{2.7}) follows from (\ref{2.9}),
\begin{eqnarray}
\Phi_{{\textrm {in}}}= {\textrm{exp}}\Big[\frac{i}{\hbar}(1+\sum_i\beta_i\frac{\hbar^i}{H_{\textrm {KN}}^i})\Big(-\omega t  -\omega\int_C\frac{dr}{\sqrt{f(r)g(r)}}\Big)\Big]
\label{2.19}
\end{eqnarray} 
and
\begin{eqnarray}
\Phi_{{\textrm {out}}}= {\textrm{exp}}\Big[\frac{i}{\hbar}(1+\sum_i\beta_i\frac{\hbar^i}{H_{\textrm {KN}}^i})\Big(-\omega t  +\omega\int_C\frac{dr}{\sqrt{f(r)g(r)}}\Big)\Big].
\label{2.20}
\end{eqnarray} 
The paths for the ingoing and outgoing particle crossing the event horizon are not same. The ingoing particle can cross the event horizon classically, whereas, the outgoing particle trajectory is classically forbidden. The metric coefficients for `$r-t$' sector alter sign at the two sides of the event horizon. Therefore, the path in which tunneling takes place has an imaginary time coordinate (${\textrm {Im}}~t$). The ingoing and outgoing probabilities are now given by,
\begin{eqnarray}
P_{{\textrm{in}}}=|\Phi_{{\textrm {in}}}|^2= {\textrm{exp}}\Big[-\frac{2}{\hbar}(1+\sum_i\beta_i\frac{\hbar^i}{H_{\textrm {KN}}^i})\Big(-\omega{\textrm{Im}}~t -\omega{\textrm{Im}}\int_C\frac{dr}{\sqrt{f(r)g(r)}}\Big)\Big]
\label{2.21}
\end{eqnarray}
and
\begin{eqnarray}
P_{{\textrm{out}}}=|\Phi_{{\textrm {out}}}|^2= {\textrm{exp}}\Big[-\frac{2}{\hbar}(1+\sum_i\beta_i\frac{\hbar^i}{H_{\textrm {KN}}^i})\Big(-\omega{\textrm{Im}}~t +\omega{\textrm{Im}}\int_C\frac{dr}{\sqrt{f(r)g(r)}}\Big)\Big].
\label{2.22}
\end{eqnarray}
Since in the classical limit ($\hbar\rightarrow 0$) $P_{{\textrm {in}}}$ is unity, one has,
\begin{eqnarray}
{\textrm{Im}}~t = -{\textrm{Im}}\int_C\frac{dr}{\sqrt{f(r)g(r)}}.
\label{2.23}
\end{eqnarray}
The presence of this imaginary time component is in agreement with \cite{Pilling, Majhiconnect}, where it is shown that for the Schwarzschild black hole if one connects the two patches (in Kruskal-Szekeres coordinates) exterior and interior to the event horizon, there is a contribution coming from the imaginary time coordinate. The value of this contribution is $2\pi i M$ which exactly coincides with (\ref{2.23}) evaluated for the Schwarzschild case with $f(r)= g(r)= (1- \frac{2M}{r})$ \cite{Majhiconnect}.

As a result the outgoing probability for the tunneling particle becomes,
\begin{eqnarray}
P_{{\textrm{out}}}={\textrm{exp}}\Big[-\frac{4}{\hbar}\omega\Big(1+\sum_i\beta_i\frac{\hbar^i}{H_{\textrm {KN}}^i}\Big){\textrm{Im}}\int_C\frac{dr}{\sqrt{f(r)g(r)}}\Big].
\label{2.24}
\end{eqnarray}
The principle of ``detailed balance'' \cite{Paddy} for the ingoing and outgoing probabilities states that,
\begin{eqnarray}
P_{{\textrm{out}}}= {\textrm {exp}}\Big(-\frac{\omega}{T_{\textrm {bh}}}\Big)P_{\textrm{in}}={\textrm{exp}} \Big(-\frac{\omega}{T_{\textrm bh}}\Big)
\label{2.25}
\end{eqnarray}
Comparing (\ref{2.24}) and (\ref{2.25}) the corrected Hawking temperature for the Kerr-Newman black hole is given by
\begin{eqnarray}
T_{\textrm {bh}}=T_{\textrm H}\Big(1+\sum_i\beta_i\frac{\hbar^i}{H_{\textrm {KN}}^i}\Big)^{-1},
\label{2.26}
\end{eqnarray}
where
\begin{eqnarray}
T_{\textrm H} = \frac{\hbar}{4}\Big({\textrm{Im}}\int_C\frac{dr}{\sqrt{f(r)g(r)}}\Big)^{-1}
\label{2.27}
\end{eqnarray}
is the semiclassical Hawking temperature. Using the expressions of $f(r)$ and $g(r)$ form (\ref{2.71}) it follows that,
\begin{eqnarray}
T_{\textrm H} = \frac{\hbar\sqrt{F'(r_+,\theta_0)g'(r_+,\theta_0)}}{4\pi}=\frac{\hbar}{2\pi}\frac{(r_+ -M)}{(r^2_+ +a^2)},
\label{2.28}
\end{eqnarray}
 which is the familiar result for the semiclassical Hawking temperature for the Kerr-Newman black hole.

\subsection{Fermion tunneling}


In this section we discuss Hawking effect through the tunneling of fermions. Although a reasonable literature exists for the computation of the semiclassical Hawking temperature \cite{fermion, fermoth, Majhifermion}, there is no analysis on possible corrections, for a general metric, within this framework. There is a paper \cite{Majhifermion} which discusses such corrections but only for the Schwarzschild metric. Here we shall do the analysis for the tunneling of massless fermions from the Kerr-Newman spacetime and reproduce the expressions (\ref{2.26}) and (\ref{2.28}) which were obtained for a scalar particle tunneling.

The Dirac equation for massless fermions is given by
\begin{eqnarray}
i\gamma ^{\mu }D_{\mu }\psi  =0,
\label{3.1}
\end{eqnarray} 
 where the covariant derivative is defined as,
\begin{eqnarray}
D_{\mu } =\partial _{\mu }+\frac{1}{2}i\Gamma _{\text{ \ }\mu }^{\alpha \text{ \ }%
\beta }\Sigma _{\alpha \beta }\nonumber\\
\Gamma^{\alpha~\beta}_{~\mu}= g^{\beta\nu}\Gamma^{\alpha}_{\mu\nu}
\label{3.2}
\end{eqnarray} 
and
\begin{eqnarray}
\Sigma _{\alpha \beta } =\frac{1}{4}i[\gamma _{\alpha },\gamma _{\beta }]
\label{3.3}
\end{eqnarray} 
The $\gamma^{\mu }$ matrices satisfy the anticommutation relation $\{\gamma ^{\mu },\gamma ^{\nu}\}=2g^{\mu \nu }\times {\bf 1}$.

We are concerned only with the radial trajectory and for this it is useful to work with the metric (\ref{2.7}). Using this one can write (\ref{3.1}) as
\begin{eqnarray}
i\gamma^{\mu}\partial_{\mu}\psi- \frac{1}{2}\left(g^{tt}\gamma^{\mu}\Gamma_{\mu t}^{r}-g^{rr}\gamma^{\mu}\Gamma_{\mu r}^{t}\right)\Sigma_{rt}\psi=0
\label{3.4}
\end{eqnarray}
The nonvanishing connections which contribute to the resulting equation are

\begin{eqnarray}
\Gamma^{r}_{tt}= \frac{f'g}{2}; \Gamma^{t}_{tr}= \frac{f'}{2f}.
\label{3.5}
\end{eqnarray}
Let us define the $\gamma$ matrices for the `$r-t$' sector as

\begin{eqnarray}
\gamma ^{t} &=&\frac{1}{\sqrt{f(r)}}\gamma ^{0},~~~~~\gamma^{r}=\sqrt{g(r)}\gamma ^{3},
\label{3.6}
\end{eqnarray}
where $\gamma^0$ and $\gamma^3$ are members of the standard Weyl or chiral representation of $\gamma$ matrices \cite{fermion} in Minkwoski spacetime, expressed as

\begin{eqnarray}
\gamma ^{0} &=&\left( 
\begin{array}{cc}
0 & I \\ 
-I & 0%
\end{array}%
\right) \
 \text{ \ \ \ }\gamma ^{3}=\left( 
\begin{array}{cc}
0 & \sigma ^{3} \\ 
\sigma ^{3} & 0%
\end{array}%
\right).
\label{3.7}
\end{eqnarray}

Using (\ref{3.3}), (\ref{3.5}) and (\ref{3.6}) the equation of motion (\ref{3.4}) is simplified as,
\begin{eqnarray}
i\gamma^t\partial_t\psi+i\gamma^r\partial_r\psi+\frac{f'(r)g(r)}{2f(r)}\gamma^{t}\Sigma_{rt}\psi=0,
\label{3.8}
\end{eqnarray}
where 
\begin{eqnarray}
\Sigma_{rt}=\frac{i}{2}\sqrt\frac{f(r)}{g(r)}{\left(\begin{array}{c c c c}
1 & 0 & 0 & 0\\
0 & -1 & 0 & 0\\
0 & 0 & -1 & 0\\
0 & 0 & 0 & 1
\end{array}\right).}
\label{3.9}
\end{eqnarray}

The spin up (+ ve `$r$' direction) and spin down (- ve `$r$' direction) ansatz for the Dirac field have the following forms respectively,  
\begin{eqnarray}
\psi_\uparrow(t,r)= \left(\begin{array}{c}
A(t,r) \\
0 \\
B(t,r) \\
0
\end{array}\right){\textrm{exp}}\Big[\frac{i}{\hbar}I_\uparrow (t,r)\Big]
\label{3.91}
\end{eqnarray}  
and
\begin{eqnarray}
\psi_\downarrow(t,r)= \left(\begin{array}{c}
0 \\
C(t,r) \\
0 \\
D(t,r) \\
\end{array}\right){\textrm{exp}}\Big[\frac{i}{\hbar}I_\downarrow (t,r)\Big].
\label{3.10}
\end{eqnarray}
 Here $I_{\uparrow}(r,t)$ is the action for the spin up case and will be expanded in powers of $\hbar$. We shall perform our analysis only for the spin up case since the spin down case is fully analogous. On substitution of the ansatz (\ref{3.91}) in (\ref{3.8}) and simplifying, we get the following two nonzero equations,
\begin{eqnarray}
B(t,r)[\partial_t{I_{\uparrow}(r,t)}+\sqrt{fg}\partial_rI_{\uparrow}(r,t)]=0
\label{3.11}
\end{eqnarray}
and
\begin{eqnarray}
A(t,r)[\partial_t{I_{\uparrow}(r,t)}-\sqrt{fg}\partial_rI_{\uparrow}(r,t)]=0.
\label{3.12}
\end{eqnarray}
Now let us expand all the variables in the `$r-t$' sector in powers of $\hbar$, as
\begin{eqnarray}
I_{\uparrow}(r,t)=I(r,t)=I_0(r,t)+\displaystyle\sum_i \hbar^i I_i(r,t)\nonumber\\
A(r,t)=A_0(r,t)+\displaystyle\sum_i \hbar^i A_i(r,t)\label{3.13}\\
B(r,t)=B_0(r,t)+\displaystyle\sum_i \hbar^i B_i(r,t).\nonumber
\end{eqnarray}
Substituting all the terms from (\ref{3.13}) into (\ref{3.11}) and (\ref{3.12}) yields  (for $a= 0, 1, 3...$)
\begin{eqnarray}
B_a(r,t)\left(\partial_t I_a(r,t)+\sqrt{fg}~\partial_rI_a(r,t)\right)=0\nonumber\\
A_a(r,t)\left(\partial_t I_a(r,t)-\sqrt{fg}~\partial_rI_a(r,t)\right)=0.
\label{3.14}
\end{eqnarray}
Thus we have the following sets of solutions, respectively, for $B_a{\textrm {'s}}\neq0$ and $A_a{\textrm {'s}}\neq0$,
\begin{eqnarray}
{\textrm{Set-I}:}~~~ \partial_t I_a(r,t)+\sqrt{fg}~\partial_rI_a(r,t)=0   
\label{3.15}
\end{eqnarray}
\begin{eqnarray}
{\textrm{Set-II}:}~~~ \partial_t I_a(r,t)-\sqrt{fg}~\partial_rI_a(r,t)=0.   
\label{3.16}
\end{eqnarray}
Similar to the scalar particle tunneling here also one can separate the semiclassical action for the `$r-t$' sector as
\begin{eqnarray}
I_0(r,t)=-\omega t +W_0(r),
\label{3.17}
\end{eqnarray}
 where $\omega=(E-P_{\phi}\Omega_{\textrm H})$. Substituting (\ref{3.17}) in (\ref{3.15}) and (\ref{3.16}) for ($a=0$) and integrating we get
\begin{eqnarray}
W_0^{\pm}(r)=\pm\omega\int_C\frac{dr}{\sqrt{f(r)g(r)}}
\label{3.18}
\end{eqnarray}
and subsequently
\begin{eqnarray}
I_0(r,t)=\left(-\omega t \pm\omega\int_C\frac{dr}{\sqrt{f(r)g(r)}}\right),
\label{3.19}
\end{eqnarray}
where + (-) sign implies that the particle is outgoing (ingoing). Because of the similar structure of (\ref{3.15}) and (\ref{3.16}), for all ($a=0, 1, 2... $), the solutions for $I_i(r,t)$' s can at most differ by a proportionality factor from $I_0(r,t)$ and the most general solution for $I(r,t)$ is given by
\begin{eqnarray}
I(r,t)=(1+{\displaystyle\sum_i{\gamma_i\hbar^i}})\left(-\omega t \pm\omega\int_C\frac{dr}{\sqrt{f(r)g(r)}}\right).
\label{3.191}
\end{eqnarray}
This is an exact analogue of the scalar particle tunneling case (\ref{2.18}) and one can check this will lead to an identical expression of corrected Hawking temperature as given by (\ref{2.26}) and (\ref{2.28}) by exactly mimicking the steps discussed there.

\section{Exact differential and Corrected Area Law}
With the result of corrected Hawking temperature (\ref{2.26}) we now proceed with the calculation of the corrected entropy and area law. The modified form of first law of thermodynamics for Kerr-Newman black hole in the presence of corrections to Hawking temperature is 
\begin{eqnarray}
dS_{\textrm {bh}}=\frac{dM}{T_{\textrm {bh}}}+(-\frac{\Omega_{\textrm H}}{T_{\textrm {bh}}})dJ+(-\frac{\Phi_{\textrm H}}{T_{\textrm {bh}}})dQ.
\label{5.1}
\end{eqnarray}
 In this context we want to stress that, 
\begin{itemize}
\item{{\it Entropy must be a state function for all stationary spacetimes even in the presence of the quantum corrections to the semiclassical value.}}
\end{itemize}
 This implies that $dS_{\textrm {bh}}$ has to be an exact differential. In the expression for $T_{\textrm {bh}}$ in (\ref{2.26}) there are six undetermined coefficients ($a_1$ to $a_6$) present in $H_{{\textrm {KN}}}$ (\ref{2.15}). The first step in the analysis is to fix these coefficients in such a way that $dS_{\textrm {bh}}$ in (\ref{5.1}) remains an exact deferential. By this restriction we make the corrected black hole entropy independent of any collapse process. For (\ref{5.1}) to be an exact differential the following relations must hold:
\begin{eqnarray}
\frac{\partial}{\partial J}(\frac{1}{T_{\textrm {bh}}})\big|_{M,Q}=\frac{\partial}{\partial M}(-\frac{\Omega_{\textrm H}}{T_{\textrm {bh}}})\big|_{J,Q}
\label{5.2}
\end{eqnarray}
\begin{eqnarray}
\frac{\partial}{\partial Q}(-\frac{\Omega_{\textrm H}}{T_{\textrm {bh}}})\big|_{M,J}=\frac{\partial}{\partial J}(-\frac{\Phi_{\textrm H}}{T_{\textrm {bh}}})\big|_{M,Q}
\label{5.3}
\end{eqnarray}
\begin{eqnarray}
\frac{\partial}{\partial M}(-\frac{\Phi_{\textrm H}}{T_{\textrm {bh}}})\big|_{J,Q}=\frac{\partial}{\partial Q}(\frac{1}{T_{\textrm {bh}}})\big|_{J,M}.
\label{5.4}
\end{eqnarray}
Using the expression of $T_{\textrm {bh}}$ from (\ref{2.26}) and the semiclassical result from (\ref{4.71}), the first condition (\ref{5.2}) reduces to 
\begin{eqnarray}
\frac{\partial}{\partial J}\displaystyle\sum_i \frac{\beta_i\hbar^i}{H^i_{\textrm {KN}}}\big|_{M,Q}=-\Omega_{\textrm H}\frac{\partial}{\partial M}\displaystyle\sum_i\frac{\beta_i\hbar^i}{H^i_{\textrm {KN}}}\big|_{J,Q}
\label{5.41}
\end{eqnarray}
Expanding this equation in powers of $\hbar$, one has the following equality
\begin{eqnarray}
\frac{\partial H_{{\textrm {KN}}}}{\partial J}\big|_{M,Q}=-\Omega_{\textrm H}\frac{\partial H_{{\textrm {KN}}}}{\partial M}\big|_{J,Q}.
\label{5.5}   
\end{eqnarray}
Similarly the other two integrability conditions (\ref{5.3}) and (\ref{5.4}) lead to other conditions on $H_{\textrm {KN}}$,
\begin{eqnarray}
\frac{\partial H_{{\textrm {KN}}}}{\partial Q}\big|_{M,J}=\left(\frac{\Phi_{\textrm H}}{\Omega_{\textrm H}}\right)\frac{\partial{H_{\textrm {KN}}}}{\partial J}\big|_{M,Q}
\label{5.6}   
\end{eqnarray}

\begin{eqnarray}
\frac{\partial H_{{\textrm {KN}}}}{\partial M}\big|_{J,Q}=-\frac{1}{\Phi_{\textrm H}}\frac{\partial H_{\textrm {KN}}}{\partial Q}\big|_{J,M}
\label{5.61}   
\end{eqnarray}
respectively. The number of unknown coefficients present in $H_{{\textrm {KN}}}$ is six and we have only three equations involving them, so the problem is under determined.

 As a remedy to this problem let us first carry out the dimensional analysis for Kerr spacetime and then use the result to reduce the arbitrariness in $H_{\textrm {KN}}$. For $Q=0$ the Kerr-Newman metric reduces to the rotating Kerr spacetime and one can carry the same analysis to find the corrections to Hawking temperature for both scalar particle and fermion tunneling from Kerr spacetime. An identical calculation will be repeated with $Q=0$. The only difference will appear in the dimensional analysis (\ref{2.15}). Since Kerr metric is chargeless the most general expression for corrected Hawking temperature will come out as
\begin{eqnarray}
T_{\textrm {bh}}=T\Big(1+\sum_i\beta_i\frac{\hbar^i}{H_{\textrm K}^i}\Big)^{-1},
\label{5.7}
\end{eqnarray}
where $H_{\textrm K}$ is now given by
\begin{eqnarray}
H_{{\textrm K}}=H_{\textrm {KN}}(Q=0)= a_1r^2_+ + a_2Mr_++a_3M^2.
\label{5.8}
\end{eqnarray}
The first law of thermodynamics for Kerr black hole is
\begin{eqnarray}
dS=\frac{dM}{T_{\textrm H}}+(-\frac{\Omega_{\textrm H}}{T_{\textrm H}})dJ,
\label{5.9}
\end{eqnarray}
where $T_{\textrm H}$ and $\Omega_{\textrm H}$ for Kerr black hole are obtained from their corresponding expressions for the Kerr-Newman case, for $Q=0$, as given in Appendix 7.1. With these expressions one can easily check that $dS$ is an exact differential for Kerr black hole as well since the only integrability condition 
\begin{eqnarray}
\frac{\partial}{\partial J}(\frac{1}{T_{\textrm H}})\big|_{M}=\frac{\partial}{\partial M}(-\frac{\Omega_{\textrm H}}{T_{\textrm H}})\big|_{J}
\label{5.10}
\end{eqnarray} 
is satisfied. As stated earlier the idea behind introducing the Kerr spacetime is to carry out the dimensional analysis for Kerr spacetime first, then demanding that for $Q=0$ the dimensional parameter $H_{{\textrm {KN}}}$ will be same as $H_{{\textrm K}}$. The form of first law for Kerr black hole in presence of corrections to the Hawking temperature, is given by  
\begin{eqnarray}
dS_{\textrm {bh}}=\frac{dM}{T_{\textrm {bh}}}+(-\frac{\Omega_{\textrm H}}{T_{\textrm {bh}}})dJ,
\label{5.11}
\end{eqnarray}
 where the general form of $T_{\textrm {bh}}$ is given in (\ref{5.7}). Now demanding that the corrected entropy of Kerr black hole must be a state function, the following integrability condition  
\begin{eqnarray}
\frac{\partial}{\partial J}(\frac{1}{T_{\textrm {bh}}})\big|_{M}=\frac{\partial}{\partial M}(-\frac{\Omega_{\textrm H}}{T_{\textrm {bh}}})\big|_{J}
\label{5.12}
\end{eqnarray} 
 must hold. Using the semiclassical result from (\ref{5.10}) and considering corrections to all orders in $\hbar$ to the Hawking temperature in (\ref{5.7}) it follows that the above integrability condition is satisfied if the following relation holds
\begin{eqnarray}
\frac{\partial H_{{\textrm K}}}{\partial J}\big|_{M}=-\Omega_{\textrm H}\frac{\partial H_{{\textrm K}}}{\partial M}\big|_{J}.
\label{5.13}
\end{eqnarray} 
 From (\ref{5.8}) it follows that this equality holds only for
\begin{eqnarray}
a_1= 0= a_3
\label{5.131}
\end{eqnarray}
 and the form of $H_{\textrm K}$ is given by 
\begin{eqnarray}
H_{{\textrm K}}=a_2Mr_+.
\label{5.14}
\end{eqnarray} 
Therefore, the corrected form for the Hawking temperature obeying the integrability condition (\ref{5.12}) for the Kerr black hole is given by
\begin{eqnarray}
T_{\textrm {bh}}=T_{\textrm H}\left(1+\displaystyle\sum_i\frac{\beta_i\hbar^i}{(a_2Mr_+)^i}\right)^{-1}=T_{\textrm H}\left(1+\displaystyle\sum_i\frac{\tilde\beta_i\hbar^i}{(Mr_+)^i}\right)^{-1}.
\label{5.141}
\end{eqnarray}

The natural expectation from the dimensional term ($H_{\textrm {KN}}$) in (\ref{2.15}) is that for $Q=0$ it gives the correct dimensional term ($H_{\textrm K}$) in (\ref{5.14}). To fulfil this criterion we must have $a_1= 0= a_3$ in (\ref{2.15}) and this leads to
\begin{align}
H_{{\textrm {KN}}}=a_2Mr_+ + a_4{r_+ Q}+a_5{MQ}+a_6{Q^2}\notag\\
=a_2(Mr_+ + \tilde a_4{r_+ Q}+\tilde a_5{MQ}+\tilde a_6{Q^2}),
\label{5.15}
\end{align}
where $\tilde a_j=\frac{a_j}{a_2}$. Now we are in a position to find the precise form of the dimensional term ($H_{\textrm {KN}}$) satisfying the integrability conditions given in (\ref{5.5}), (\ref{5.6}) and (\ref{5.61}). Note that with the modified expression (\ref{5.15}) the problem of under determination of six coefficients by only three integrability conditions for Kerr-Newman spacetime has been removed. With this expression of $H_{\textrm {KN}}$ one has effectively three undetermined coefficients with three equations and it is straightforward to calculate those coefficients. Putting the new expression of $H_{\text {KN}}$ in (\ref{5.5}), (\ref{5.6}) and (\ref{5.61}) one obtains,      
\begin{eqnarray}
\tilde a_5 - \tilde a_4\frac{r_+}{M}=0 
\label{5.16}
\end{eqnarray}
\begin{eqnarray}
2\tilde a_6 Q + \tilde a_4 \left(\frac{Mr_+ +Q^2}{M}\right)+ \tilde a_5M=-Q 
\label{5.17}
\end{eqnarray}
\begin{eqnarray}
2\tilde a_6 Q + \tilde a_4 \left(r_+ + \frac{J^2Q^2}{M^3(r^2_+ + J^2/M^2)}\right)+ \tilde a_5\left(M+\frac{Q^2r_+}{r^2_+ +J^2/M^2}\right)=-Q. 
\label{5.171} 
\end{eqnarray}
The simultaneous solution of these three equations yields,
\begin{eqnarray}
\tilde a_4= 0= \tilde a_5\notag\\
\tilde a_6= -\frac{1}{2}.
\label{5.172}
\end{eqnarray}
As a result the final form of $H_{\text {KN}}$ derived by the requirements:\\ (i) $H_{\text {KN}}$  must satisfy the integrability conditions (\ref{5.5}, \ref{5.6}, \ref{5.61}), \\ (ii)   $H_{\text {KN}}= H_{\text K}$ for $Q=0$,\\  is given by 
\begin{eqnarray}
H_{{\textrm {KN}}}=a_2(Mr_+-\frac{1}{2}Q^2).
\label{5.18}
\end{eqnarray}
Hence the corrected Hawking temperature for Kerr-Newman black hole is found to be
\begin{eqnarray}
T_{\textrm {bh}}=T\left(1+\displaystyle\sum_i\frac{\beta_i\hbar^i}{a^i_2(Mr_+-\frac{Q^2}{2})^i}\right)^{-1}=T\left(1+\displaystyle\sum_i\frac{\tilde\beta_i\hbar^i}{(Mr_+-\frac{Q^2}{2})^i}\right)^{-1},
\label{5.19}
\end{eqnarray} 
where $\tilde\beta_i=\frac{\beta_i}{a^i_2}$.

We are now in a position to compute the corrected entropy and find the deviations from the semiclassical area law. Comparing (\ref{4.8}) and (\ref{5.1}) with $T_{\textrm {bh}}$ given above we find a similar dictionary as (\ref{4.15}) by modifying semiclassical terms with corrected versions, where necessary, as 
\begin{align}
(f\rightarrow S_{\textrm {bh}},~~x\rightarrow M,~~y\rightarrow J,~~z\rightarrow Q)\nonumber\\
(U\rightarrow\frac{1}{T_{\textrm {bh}}},~~V\rightarrow\frac{-\Omega_{\textrm H}}{T_{\textrm {bh}}},~~W\rightarrow\frac{-\Phi_{\textrm H}}{T_{\textrm {bh}}}).
\label{5.20}
\end{align}
Following this dictionary and (\ref{4.12}), (\ref{4.13}) and (\ref{4.14}) the corrected entropy for Kerr-Newman black hole has the form
\begin{eqnarray}
S_{\textrm {bh}}=\int{\frac{dM}{T_{\textrm {bh}}}}+\int{XdJ}+\int{YdQ},
\label{5.21}
\end{eqnarray}
where
\begin{eqnarray}
X=(-\frac{\Omega_{\textrm H}}{T_{\textrm {bh}}})-\frac{\partial}{\partial J}{\int\frac{dM}{T_{\textrm {bh}}}}
\label{5.22}
\end{eqnarray}
and
\begin{eqnarray}
Y=(-\frac{\Phi_{\textrm H}}{T_{\textrm {bh}}})-\frac{\partial}{\partial Q}[{\int\frac{dM}{T_{\textrm {bh}}}}+{\int XdJ}].
\label{5.23}
\end{eqnarray}
It is possible to calculate $S_{\textrm {bh}}$ analytically up to all orders of $\hbar$. However we shall restrict ourselves up to second order correction to the Hawking temperature. Integration over $M$ yields,   
\begin{eqnarray}
\int\frac{dM}{T}=\frac{\pi}{\hbar}(2Mr_+-Q^2)+ 2\pi\tilde\beta_1\hbar\log(2Mr_+-Q^2)-\frac{4\pi\tilde\beta_2\hbar^2}{(2Mr_+-Q^2)^2}+{\textrm {const.}}+{\textrm {higher order terms}}.
\label{5.24}
\end{eqnarray}
With this result of integration one can check the following relation, 
\begin{eqnarray}
\frac{\partial}{\partial J}\int\frac{dM}{T_{\textrm {bh}}}=-\frac{\Omega_{\textrm H}}{T_{\textrm {bh}}}.
\label{5.25}
\end{eqnarray}
 Therefore $X=0$. Furthermore we get
\begin{eqnarray}
\frac{\partial}{\partial Q}\int\frac{dM}{T_{\textrm {bh}}}=-\frac{\Phi_{\textrm H}}{T_{\textrm {bh}}},
\label{5.26}
\end{eqnarray}
 and using this equality together with $X=0$ we find $Y=0$. The fact that both $X$ and $Y$ pick the most trivial solution as zero in the black hole context, both with or without quantum corrections, is quite unique. The final result for the entropy of the Kerr-Newman black hole in presence of quantum corrections is now given by
\begin{eqnarray}
S_{\textrm {bh}}=\frac{\pi}{\hbar}(2Mr_+-Q^2)+ 2\pi\tilde\beta_1\log(2Mr_+-Q^2)-\frac{4\pi\tilde\beta_2\hbar}{(2Mr_+-Q^2)}+{\textrm {const.}}+{\textrm {higher order terms}}.
\label{5.27}
\end{eqnarray}  
In terms of the semiclassical black hole entropy and horizon area this can be expressed, respectively, as
\begin{eqnarray}
S_{\textrm {bh}}=S_{\textrm {BH}}+ 2\pi\tilde\beta_1\log S_{\textrm {BH}}-\frac{4\pi^2\tilde\beta_2}{S_{\textrm {BH}}}+{\textrm {const.}}+{\textrm {higher order terms}}.
\label{5.28}
\end{eqnarray}  
and
\begin{eqnarray}
S_{\textrm {bh}}=\frac{A}{4}+ 2\pi\tilde\beta_1\log A-\frac{16\pi^2\tilde\beta_2}{A}+{\textrm {const.}}+{\textrm {higher order terms}}.
\label{5.29}
\end{eqnarray}  
The first term in the expression (\ref{5.28}) is the usual semiclassical Bekenstein-Hawking entropy and the other terms are due to quantum corrections. The logarithmic and inverse area terms have appeared as the leading and non leading corrections to the entropy and area law.

\section{Determination of the leading correction to entropy by trace anomaly}

In the expression for entropy in (\ref{5.29}) the leading order correction includes an arbitrary coefficient $\tilde\beta_1$. In this section we shall determine this coefficient by using trace anomaly.   

Consider the scalar particle tunneling case. The expression for the action for the Kerr-Newman spacetime is given by (\ref{2.16})
\begin{eqnarray}
{\cal S}(r,t)=\left({\cal S}_0(r,t)+\sum_i\hbar^i{\cal S}_i(r,t)\right)=\left({\cal S}_0(r,t)+\sum_i\frac{\tilde\beta_i\hbar^i}{(Mr_+ -\frac{Q^2}{2})^i}{\cal S}_0(r,t)\right),
\label{6.1}
\end{eqnarray}
 where the appropriate form for $H_{KN}$ from (\ref{5.18}) is considered. Taking the first order ($\hbar$) correction in this equation we can write the following relation for the imaginary part of the outgoing particle action   
\begin{eqnarray}
{\textrm {Im}{\cal S}_1^{\textrm {out}}}(r,t)=\frac{\tilde\beta_1}{(Mr_+ -\frac{Q^2}{2})}{\textrm {Im}}{\cal S}_0^{\textrm {out}}(r,t).
\label{6.2}
\end{eqnarray}
The imaginary part for the semiclassical action for an outgoing particle can be found from (\ref{2.13}), (\ref{2.17}) and (\ref{2.23}) as 
\begin{eqnarray}
{\textrm {Im}}{\cal S}_0^{\textrm {out}}(r,t)=-2\omega {\textrm {Im}}\int_C{\frac{dr}{\sqrt{f(r)g(r)}}}.
\label{6.3}
\end{eqnarray}

Let us make an infinitesimal scale transformation of the metric coefficients in (\ref{2.7}) parametrized by the constant factor `$k$' \cite{Hawkzeta, Majhitrace} such that $\bar f(r)=k f(r)\simeq (1+\delta k)f(r)$ and $\bar g(r)=k^{-1} g(r)\simeq (1+\delta k)^{-1}g(r)$. From the scale invariance of the Klein-Gordon equation in (\ref{2.8}) it follows that the Klein-Gordon field ($\Phi$) should transform as $\Phi=k^{-1}\Phi$. Since $\Phi$ has a dimension of mass, one interprets that the black hole mass ($M$) should transform as $M=k^{-1}M\simeq (1+\delta k)^{-1}M$ under the infinitesimal scale transformation. Therefore the other two black hole parameters ($Q$, $a$) and the particle energy $\omega$ should also transform as $M$ does. Using these it is straightforward to calculate the transformed form of (\ref{6.2}) and (\ref{6.3}) to get

\begin{eqnarray}
{\textrm {Im}\overline{\cal S}_1^{\textrm {out}}}(r,t)=\frac{\tilde\beta_1}{(\overline M\overline r_+ -\frac{\overline Q^2}{2})}{\textrm {Im}}\overline{\cal S}_0^{\textrm {out}}(r,t)=\frac{\tilde\beta_1}{(Mr_+ -\frac{Q^2}{2})}(1+\delta k){\textrm {Im}}{\cal S}_0^{\textrm {out}}(r,t),
\label{6.4}
\end{eqnarray}
and
\begin{eqnarray}
\frac{\delta {\textrm {Im}}{{\cal S}}^{{\textrm {out}}}_1(r,t)}{\delta k}=\frac{\tilde\beta_1}{(Mr_+ -\frac{Q^2}{2})}{\textrm {Im}}{\cal S}_0^{\textrm {out}}(r,t).
\label{6.5}
\end{eqnarray} 
Now considering the scalar field Lagrangian it can be shown that under a constant scale transformation of the metric coefficients the action is not invariant in the presence of trace anomaly and this lack of conformal invariance is given by the following relation 
\begin{eqnarray}
\frac{\delta {{\cal S}}(r,t)}{\delta k}=\frac{1}{2}\int{d^4x\sqrt{-g}(<T^{\mu}_{~\mu}>^{(1)}+<T^{\mu}_{~\mu}>^{(2)}+...)},
\label{6.6}
\end{eqnarray} 
 where $<T^{\mu}_{~\mu}>^{(i)}$' s  are the trace of the regularised stress energy tensor calculated for $i$-th loop. However, in the literature \cite{Hawkzeta, Dewitt}, only the first order loop calculation has been carried out and this gives
\begin{eqnarray}
\frac{\delta {\textrm {Im}}{{\cal S}}^{{\textrm {out}}}_1(r,t)}{\delta k}=\frac{1}{2}{\textrm {Im}}\int{d^4x\sqrt{-g}(<T^{\mu}_{~\mu}>^{(1)})},
\label{6.7}
\end{eqnarray} 
where, for a scalar background, the form of trace anomaly is given by 
\begin{eqnarray}
<T^{\mu}_{~\mu}>^{(1)}=\frac{1}{2880\pi^2}\left(R_{\mu\nu\rho\sigma}R^{\mu\nu\rho\sigma}-R_{\mu\nu}R^{\mu\nu}+\nabla_{\mu}\nabla^{\mu}R\right)
\label{6.9}
\end{eqnarray} 
Now integrating (\ref{6.3}) around the pole at $r=r_+$ we get
\begin{eqnarray}
{\textrm {Im}}{\cal S}_0^{\textrm {out}}(r,t)=-2\pi\omega\frac{(r_+M-\frac{Q^2}{2})}{(r_+-M)}
\label{6.9a}
\end{eqnarray}
 Now putting this in (\ref{6.5}) and comparing with (\ref{6.7}) we find
\begin{eqnarray}
\tilde\beta_1=-\frac{(M^2-Q^2-\frac{J^2}{M^2})^{1/2}}{4\pi\omega}{\textrm {Im}}\int{d^4x\sqrt{-g}<T^{\mu}_{~\mu}>^{(1)}}.
\label{6.8}
\end{eqnarray} 
Equation (\ref{6.8}) gives the general form of the coefficient associated with the leading correction to the semiclassical entropy for any stationary black hole. To get $\tilde\beta_1$ for a particular black hole in (3+1) dimensions one needs to solve both (\ref{6.8}) and (\ref{6.9}) for that black hole. Now we shall take different spacetime metrics and explicitly calculate $\tilde\beta_1$ for them.

\subsection{Schwarzschild Black Hole}
For $Q=0=J$ the Kerr-Newman spacetime metric reduces to the Schwarzschild spacetime and from (\ref{6.8}) it follows that
\begin{eqnarray}
\tilde\beta_1=-\frac{M}{4\pi\omega}{\textrm {Im}}\int{d^4x\sqrt{-g}<T^{\mu}_{~\mu}>^{(1)}}.
\label{6.10}
\end{eqnarray} 
The identification of particle energy by (\ref{2.13a}) is now given by $\omega=E$, where `$E$' is the Komar conserved quantity corresponding the timelike Killing vector $\frac{\partial}{\partial t}$ for the spherically symmetric Schwarzschild spacetime. An exact calculation \cite{Carrol} of the Komar integral gives $E=M$ ({\it see} Appendix 7.2), where $M$ is the mass of Schwarzschild black hole. Therefore we get 
\begin{eqnarray}
\tilde\beta_1=-\frac{1}{4\pi}{\textrm {Im}}\int{d^4x\sqrt{-g}<T^{\mu}_{~\mu}>^{(1)}}.
\label{6.11}
\end{eqnarray} 
A similar result was found by Hawking \cite{Hawkzeta}, where the path integral approach based on zeta function regularization was adopted. The path integral for standard Einstein-Hilbert gravity  was modified due to the fluctuations coming from the scalar field in the black hole spacetime. 

To find the trace anomaly of the stress tensor (\ref{6.9}) we calculate the following invariant scalars for Schwarzschild black hole, given by 
\begin{align}
R_{\mu\nu\rho\sigma}R^{\mu\nu\rho\sigma} &=\frac{48M^2}{r^6},\notag\\
R_{\mu\nu}R^{\mu\nu} &=0\label{6.11a}\\
R=0.\notag
\end{align}
Using these we can find $<T^{\mu}_{~\mu}>^{(1)}$ from (\ref{6.9}) and inserting it in (\ref{6.11}) yields,
\begin{align}
\tilde\beta_1^{({\textrm {Sch}})} &=-\frac{1}{4\pi}\frac{1}{2880\pi^2}{\textrm {Im}}\int_{r=2M}^{\infty}\int_{\theta=0}^{\pi}\int_{\phi=0}^{2\pi}\int_{t=0}^{-8\pi i M}{\frac{48M^2}{r^6}r^2 \sin\theta dr d\theta d\phi dt}\notag\\
&=\frac{1}{180\pi}.
\label{6.12}
\end{align} 
The corrected entropy/area law (\ref{5.28}, \ref{5.29}) is now given by,
\begin{align}
S_{\textrm {bh}}^{\textrm {(Sch)}}=S_{\textrm {BH}}+ \frac{1}{90}\log S_{\textrm {BH}}+{\textrm {higher order terms}},
\nonumber\\
=\frac{A}{4}+\frac{1}{90}\log A + {\textrm {higher order terms.}}
\label{6.13}
\end{align}   
This reproduces the result existing in the literature \cite{Hawkzeta, Fursaev, Majhitrace}.

\subsection{Reissner-Nordstrom Black Hole}
For the Reissner-Nordstrom black hole, putting $J=0$ in (\ref{6.8}), we get
\begin{eqnarray}
\tilde\beta_1=-\frac{(M^2-Q^2)^{1/2}}{4\pi\omega}{\textrm {Im}}\int{d^4x\sqrt{-g}<T^{\mu}_{~\mu}>^{(1)}},
\label{6.14}
\end{eqnarray} 
 where the particle energy is again given by the Komar energy integral corresponding to the timelike Killing field $\frac{\partial}{\partial t}$. Unlike the Schwarzschild case, however,  the effective energy for Reissner-Nordstrom black hole observed at a distance $r$, is now given by ({\it see} Appendix 7.2),  
\begin{eqnarray} 
\omega=E=(M-\frac{Q^2}{r}).
\label{6.15} 
\end{eqnarray}   
For a particle undergoing tunneling $r=r_+=(M+\sqrt{M^2-Q^2})$, we get $\omega=(M^2-Q^2)^{1/2}$ and therefore (\ref{6.14}) gives
\begin{eqnarray}
\tilde\beta_1=-\frac{1}{4\pi}{\textrm {Im}}\int{d^4x\sqrt{-g}<T^{\mu}_{~\mu}>^{(1)}}.
\label{6.16}
\end{eqnarray}
This has exactly the same functional form as (\ref{6.11}). To calculate this integral, we first simplify the integrand given in (\ref{6.9}), for a Reissner-Nordstrom black hole,
\begin{align}
R_{\mu\nu\rho\sigma}R^{\mu\nu\rho\sigma} &=\frac{8(7Q^4-12MQ^2r+6M^2r^2)}{r^8},\notag\\
R_{\mu\nu}R^{\mu\nu} &=\frac{4Q^4}{r^8},\label{6.16a}\\
R=0.\notag
\end{align}
With these results $<T^{\mu}_{~\mu}>^{(1)}$ is obtained and, finally,
\begin{align}
\tilde\beta_1^{(\textrm {RN})} &=-\frac{1}{4\pi}\frac{1}{2880\pi^2}{\textrm {Im}}\int_{r=r_+}^{\infty}\int_{\theta=0}^{\pi}\int_{\phi=0}^{2\pi}\int_{t=0}^{-i\beta}{<T^{\mu}_{~\mu}>^{(1)}r^2 \sin\theta dr d\theta d\phi dt}\notag\\
&=\frac{1}{180\pi}(1+\frac{3}{5}\frac{r_-^2}{r_+^2-r_+r_-}).
\label{6.16b}
\end{align} 
Therefore the corrected entropy/area law for a Reissner-Nordstrom black hole is now given by ({\it see} (\ref{5.28}, \ref{5.29}))
\begin{eqnarray}
S_{\textrm {bh}}^{\textrm {(RN)}}=S_{\textrm {BH}}+ \frac{1}{90}(1+\frac{3}{5}\frac{r_-^2}{r_+^2-r_+r_-})\log S_{\textrm {BH}}+{\textrm {higher order terms}}.
\nonumber\\
=\frac{A}{4}+ \frac{1}{90}(1+\frac{3}{5}\frac{r_-^2}{r_+^2-r_+r_-})\log{A}+{\textrm {higher order terms.}}
\label{6.16c}
\end{eqnarray}
Unlike the Schwarzschild black hole here the prefactor of the logarithmic term is not a pure number. This is because the presence of charge on the outer region of the event horizon includes a contribution to the matter sector. Therefore the charge ($Q$) directly affects the dynamics of the system which in turn is related to entropy. It is interesting to see that in the extremal limit the prefactor of the logarithmic term blows up, suggesting that there cannot be a smooth limit from non-extremal to the extremal case. This is in agreement with a recent paper \cite{Carrolpaper} where it is argued that the {\it extremal limit} of the Reissner-Nordstrom black hole is different from the extremal case itself. For the extremal case the region between inner and outer horizons disappears but in the {\it extremal limit} this region no longer disappears, rather it approaches a patch of $AdS_{2}\times S^{2}$. As a result the non-extremal to extremal limit is not continuous.

\subsection{Kerr Black Hole}
The Kerr black hole is the chargeless limit of the Kerr-Newman black hole. This is an axially symmetric solution of Einstein's equation and has two Killing vectors $\frac{\partial}{\partial t}$ and $\frac{\partial}{\partial \phi}$. Therefore it has two conserved quantities corresponding to those Killing directions. None of the Killing vectors is individually time-like, but the combination $(\frac{\partial}{\partial t}+\Omega \frac{\partial}{\partial \phi})$ is time-like throughout the spacetime (outside the event horizon). This combination however cannot be treated as a Killing vector because in general $\Omega$ is not constant. At the horizon $\Omega=\Omega_{\textrm H}$ is identified as the angular velocity of the horizon and the above time-like vector becomes null. This time-like vector plays a crucial role in the process of evaluating Komar integrals.

 For a Kerr black hole (\ref{6.8}) reduces to
\begin{eqnarray}
\tilde\beta_1=-\frac{(M^2-\frac{J^2}{M^2})^{1/2}}{4\pi\omega}{\textrm {Im}}\int{d^4x\sqrt{-g}<T^{\mu}_{~\mu}>^{(1)}}
\label{6.17}
\end{eqnarray}
where $\omega=(E-\Omega_H P_{\phi})$. In the Boyer-Lindquist coordinate, the Komar integrals corresponding to the Killing vectors $\frac{\partial}{\partial t}$ and  $\frac{\partial}{\partial \phi}$ are given by $E=M$ and $P_{\phi}=2J$ respectively \cite{Katz}. Here $M$ and $J$ are respectively the mass and angular momentum of the Kerr black hole. Using these expressions together with the angular velocity ($\Omega_H$) ({\it see} Appendix 7.2) we get $\omega=(M^2-\frac{J^2}{M^2})^{1/2}$ and therefore (\ref{6.17}) becomes
\begin{eqnarray}
\tilde\beta_1=-\frac{1}{4\pi}{\textrm {Im}}\int{d^4x\sqrt{-g}<T^{\mu}_{~\mu}>^{(1)}}
\label{6.18}
\end{eqnarray}
which is exactly same as the other two previous cases. The invariant scalars for Kerr spacetime are given by
\begin{align}
R_{\mu\nu\rho\sigma}R^{\mu\nu\rho\sigma} &=-\frac{96M^2\left(\alpha_1+15\alpha_2\cos{2\theta}+6a^4(a^2-10r^2)\cos{4\theta}+a^6\cos{6\theta}\right)}{(a^2+2r^2+a^2\cos{2\theta})^6},\notag\\
\alpha_1 &=(10a^6-180a^4r^2+240a^2r^4-32r^6), \notag\\
\alpha_2 &=(a^4-16a^2r^2+16r^4)\notag\\
R_{\mu\nu}R^{\mu\nu} &=0\label{6.18a}\\
R=0,\notag
\end{align}
from which the trace $<T^{\mu}_{~\mu}>^{(1)}$ in (\ref{6.9})is obtained. Now performing the integration we get  
\begin{align}
\tilde\beta_1^{(\textrm {K})} &=-\frac{1}{4\pi}\frac{1}{2880\pi^2}{\textrm {Im}}\int_{r=r_+}^{\infty}\int_{\theta=0}^{\pi}\int_{\phi=0}^{2\pi}\int_{t=0}^{-i\beta}{<T^{\mu}_{~\mu}>^{(1)}r^2 \sin\theta dr d\theta d\phi dt}\notag\\
&=\frac{1}{180\pi}.
\label{6.18b}
\end{align} 
Therefore the corrected entropy/area law for a Kerr black hole that follows from (\ref{5.28}, \ref{5.29}) is given by, 
\begin{eqnarray}
S_{\textrm {bh}}^{\textrm {(K)}}=S_{\textrm {BH}}+ \frac{1}{90}\log S_{\textrm {BH}}+{\textrm {higher order terms}},
\nonumber\\
=\frac{A}{4}+\frac{1}{90}\log A + {\textrm {higher order terms.}}
\label{6.18c}
\end{eqnarray} 
This result is identical to the Schwarzschild black hole. This can be physically explained by the following argument. The difference between the Schwarzschild and Kerr spacetimes is due to spin($J$). Unlike charge ($Q$), which has a contribution to the matter part, spin is arising in Kerr spacetime because of one extra Killing direction corresponding to the Killing field ($\partial_\phi$). This difference is purely geometrical and has nothing to do with the dynamics of the system and as a result there is no difference between the structure of corrected entropy in these two cases.

\subsection{Kerr-Newman Black Hole}
The general expression for $\tilde\beta_1$ in (\ref{6.8}) involves the total energy of the tunneling particle, given by (\ref{2.13a}). Unlike the Kerr black hole, in this case the effective energy faced by a particle at a finite distance from the horizon is not the same as felt at infinity. Because of the presence of electric charge ($Q$) it is modified. This was also the case for Reissner-Nordstrom black hole where one extra term ($-\frac{Q^2}{r}$) arose in (\ref{6.15}) due to the charge of the black hole. However, for Kerr-Newman black hole, because of its geometric structure the calculation of the extra contributions due to charge is technically more involved. Similar conclusions hold for the other conserved quantity $P_{\phi}$. On the other hand to get the exact form of total energy of the tunneling particle one needs to calculate both $E$ and $P_{\phi}$ in a closed form. In an earlier work \cite{Cohen} the closed form of $E$ was derived from the Komar integral but the explicit closed form calculation of $P_{\phi}$ from the Komar integral is still missing. For our analysis we have calculated the Komar integrals upto leading correction to both $E$ and $P_{\phi}$ ({\it see} Appendix 7.2)
\begin{align}
E &=M-\frac{Q^2}{r}+ {\cal O}(\frac{1}{r^2})\notag\\
P_{\phi} &=2(J-\frac{2Q^2a}{3r})+ {\cal O}(\frac{1}{r^2})
\label{6.19}
\end{align}       
Putting this (with $r=r_+$) in (\ref{2.13a}) and taking upto the ${\cal O}(\frac{1}{r_+})$ we get $\omega=(M^2-Q^2-\frac{J^2}{M^2})^{1/2}$. Therefore for the leading order we find $\tilde\beta_1$ as, 
\begin{eqnarray}
\tilde\beta_1=-\frac{1}{4\pi}{\textrm {Im}}\int{d^4x\sqrt{-g}<T^{\mu}_{~\mu}>^{(1)}},
\label{6.20}
\end{eqnarray}
 which is identical to the previous expressions. For now we shall take (\ref{6.20}) and perform the integral to find $\tilde\beta_1$ for Kerr-Newman black hole.   
The invariant scalars for Kerr-Newman black holes are given by
\begin{align}
R_{\mu \nu \rho \sigma } R^{\mu \nu \rho \sigma } &= \frac{128 }{(a^2+2 r^2+a^2 \cos{2 \theta })^6}
\nonumber\\ 
 &[192 r^4 (Q^2-2m r)^2-96 r^2 (Q^2-3 m r) (Q^2-2 m r)(a^2+2 r^2+a^2 \cos{2 \theta})) \nonumber \\ 
 &+ (7 Q^2-18 m r)(Q^2-6 m r) (a^2+2 r^2+a^2 cos{2 \theta})^2-3 m^2 (a^2+2 r^2+a^2 \cos{2 \theta})^3)],\notag\\
R_{\mu\nu}R^{\mu\nu} &= \frac{64 Q^4}{\left(a^2+2 r^2+a^2 \cos{2 \theta}\right)^4},\label{6.20a}\\
R=0.\notag
\end{align}
Simplifying $<T^{\mu}_{\mu}>^{(1)}$ in (\ref{6.9}) and performing the integration in (\ref{6.20}) one finds
\begin{align}
\tilde\beta _1^{\textrm {KN}} &= \frac{r_+^2 + r_+ r_- -Q^2}{5760 \pi  r_+^4 (r_+ -r_-) (r_+ r_- -Q^2)^{5/2}}\left(\alpha_1 + \frac{r_+\sqrt{r_+r_- -Q^2}}{r_+^2 + r_+r_- - Q^2} (9Q^8-\alpha_2r_+ + \alpha_3r_+^2r_-^2)\right)\label{6.20b}\\
 &{\textrm {with,}}\notag\\
\alpha_1 &=9 Q^4 [ r_+^4\tan^{-1}{(\frac{r_+}{\sqrt{-Q^2+r_- r_+}})} +  (Q^2-r_-r_+)^2\cot^{-1}{\frac{r_+}{\sqrt{-Q^2+r_- r_+}}}]\notag\\
\alpha_2 &=6Q^6r_+-41Q^4r_+^3+32r_+^4r_-^3+2Q^2r_-(9Q^4+13q^2r_+^2+32r_+^4)\notag\\
\alpha_3 &= 9Q^4+64Q^2r_+^2+32r_+^4.\notag
\end{align}
The corrected entropy/area law now follows from (\ref{5.28}) and (\ref{5.29}),
\begin{align}
S_{\textrm {bh}}^{\textrm {(KN)}}=S_{\textrm {BH}}+ \frac{r_+^2 + r_+ r_- -Q^2}{2880  r_+^4 (r_+ -r_-) (r_+ r_- -Q^2)^{5/2}}\left(\alpha_1 + \frac{r_+\sqrt{r_+r_- -Q^2}}{r_+^2 + r_+r_- - Q^2} (9Q^8-\alpha_2r_+ + \alpha_3r_+^2r_-^2)\right)\log S_{\textrm {BH}}~\nonumber\\
 +{\textrm {higher order terms}}.\nonumber\\
=\frac{A}{4}+\frac{r_+^2 + r_+ r_- -Q^2}{2880  r_+^4 (r_+ -r_-) (r_+ r_- -Q^2)^{5/2}}\left(\alpha_1 + \frac{r_+\sqrt{r_+r_- -Q^2}}{r_+^2 + r_+r_- - Q^2} (9Q^8-\alpha_2r_+ + \alpha_3r_+^2r_-^2)\right)\log A\nonumber\\
+{\textrm {higher order terms}}.
\label{6.20c}
\end{align} 
For $Q=0$ the above prefactor of the logarithmic term reduces to $\frac{1}{90}$, the coeffecient for the Kerr spacetime. 

\section{Conclusions}
Let us now summarise the findings in the present paper. We have given a new and simple approach to derive the ``first law of black hole thermodynamics'' from the thermodynamical perspective where one does not require the ``first law of black hole mechanics''. The key point of this derivation was the observation that ``black hole entropy'' is a {\it {state function}}. In the process we obtained some relations involving black hole entities, playing a role analogous to {\it {Maxwell's relations}}, which must hold for any stationary black hole. Based on these relations, we presented a systematic calculation of the semiclassical Bekenstein-Hawking entropy taking into account all the ``work terms on a black hole''. This approach is applicable to any stationary black hole solution. The standard semiclassical area law was reproduced. An interesting observation that has been come out of the calculation was that the work terms did not contribute to the final result of the semiclassical entropy.

To extend our method for calculating entropy in the presence of quantum corrections we first computed the corrected Hawking temperature in the tunneling mechanism. Both the tunneling of scalar particles and fermions were considered and they gave the same result for the corrected Hawking temperature. However this result involved a number of arbitrary constants. Demanding that the corrected entropy be a state function it was possible to find the appropriate form of the corrected Hawking temperature. By using this result we explicitly calculated the entropy with quantum corrections. In the process we again found that work terms on black hole did not contribute to the final result of the corrected entropy. This analysis was done for the Kerr-Newman spacetime and it was trivial to find the results for other stationary spacetimes like (i) Kerr, (ii) Reissner-Nordstrom and (iii) Schwarzschild by taking appropriate limits. It is important to note that the functional form for the corrected entropy is same for all the stationary black holes. The logarithmic and inverse area terms as leading and next to leading corrections were quite generic upto a dimensionless prefactor.

It was shown that the coefficient of the logarithmic correction was related with the trace anomaly of the stress tensor and explicit calculation of this coefficient was also done. This was a number ($\frac{1}{90}$) for both Schwarzschild and Kerr black hole. The fact that both Kerr and Schwarzschild black holes have identical corrections was explained on physical grounds (the difference between the metrics being purely geometrical and not dynamical) thereby serving as a nontrivial consistency check on our scheme. It may be noted that the factor ($\frac{1}{90}$) was also obtained (for the Schwarzschild case) in other approaches \cite{Hawkzeta, Fursaev} based on the direct evaluation of path integrals in a scalar background. For the charged spacetime (Reissner-Nordstrom and Kerr-Newman) the coefficients were not pure numbers, however in the $Q=0$ limit they reproduced the expressions for the corresponding chargeless versions.

\section{Appendix}
\subsection{Glossary of formulae for Kerr-Newman black hole}
The spacetime metric of the Kerr-Newman black hole in Boyer-Linquist coordinates ($t,~r,~\theta,~\phi$) is given by,
\begin{equation}
ds^{2} =-\tilde f(r,\theta )dt^{2}+\frac{dr^{2}}{\tilde g(r,\theta )}-2H(r,\theta
)dtd\phi +K(r,\theta )d\phi ^{2}+\Sigma (r,\theta )d\theta ^{2}
\label{2.1}  
\end{equation}
with the electromagnetic vector potential, 
$$A_{a} =-\frac{Qr}{\Sigma (r,\theta)}[(dt)_{a}-a\sin ^{2}\theta (d\phi )_{a}]$$ and, 
\begin{align}
\tilde f(r,\theta )& =\frac{\Delta (r)-a^{2}\sin ^{2}\theta }{\Sigma (r,\theta )} \\
\tilde g(r,\theta )& =\frac{\Delta (r)}{\Sigma (r,\theta )},  \notag \\
H(r,\theta )& =\frac{a\sin ^{2}\theta (r^{2}+a^{2}-\Delta (r))}{\Sigma
(r,\theta )}  \notag \\
K(r,\theta )& =\frac{(r^{2}+a^{2})^{2}-\Delta (r)a^{2}\sin ^{2}\theta }{%
\Sigma (r,\theta )}\sin ^{2}(\theta )  \notag \\
\Sigma (r,\theta )& =r^{2}+a^{2}\cos ^{2}\theta  \notag \\
\Delta (r)& =r^{2}+a^{2}+Q^{2}-2Mr  \notag\\
a=\frac{J}{M} \notag
\end{align}
The Kerr-Newman metric represents the most general class of stationary black hole solution of Einstein-Maxwell equations having all three parameters Mass $(M)$, Angular momentum $(J)$ and Charge $(Q)$. All other known stationary black hole solutions are encompassed by this three parameter solution.\\
(i)For $ Q=0$ it gives the rotating Kerr solution, (ii) $ J= 0$ leads to the Reissner-Nordstrom black hole, and (iii) for both  $ Q=0$ and $J=0$ the standard Schwarzschild solution is recovered.\\
For the non-extremal Kerr-Newman black hole the location of outer ($r_+$, event) and inner ($r_-$) horizons are given by setting $g^{rr}=0=g_{tt}$ or equivalently $\Delta =0$, which gives   
\begin{equation}
r_{\pm}=M\pm \sqrt{M^{2}-a^{2}-Q^{2}}.
\label{2.2}
\end{equation}
The angular velocity of the event horizon, which follows from the general expression of angular velocity for any rotating black hole, is given by
\begin{eqnarray}
\Omega_{\textrm H}= \Big[-\frac{g_{\phi t}}{g_{\phi \phi}}-\sqrt{{(\frac{g_{t\phi}}{g_{\phi\phi}})^2}-\frac{g_{tt}}{g_{\phi \phi}}}\Big]_{r=r_+}= \frac{a}{r^2_+ + a^2}.
\label{angv}
\end{eqnarray}
The electric potential at the event horizon is given by,
\begin{eqnarray}
\Phi_{\textrm H}= \frac{r_+ Q}{r^2_+ + a^2}.
\label{epot}
\end{eqnarray}
The area of the event horizon is given by,
\begin{eqnarray}
A= \int_{r_+} {{\sqrt{g_{\theta\theta}g_{\phi\phi}}}d\theta d\phi}= 4\pi (r^2_+ + a^2)
\label{area}
\end{eqnarray}
The semiclassical Hawking temperature in terms of surface gravity ($\kappa$) of the Kerr-Newman black hole is given by
\begin{eqnarray}
T_{\textrm H}=\frac{\hbar\kappa}{2\pi}=\frac{\hbar}{2\pi}\frac{(r_+ -M)}{(r^2_+ +a^2)}.
\label{hawktemp}
\end{eqnarray}
Using (\ref{2.2}), (\ref{angv}), (\ref{epot}) and (\ref{hawktemp}) one can find the following quantities,
\begin{eqnarray}
\frac{1}{T_{\textrm H}}=\frac{2\pi}{\hbar}\left(\frac{2M{[M+(M^2-\frac{J^2}{M^2}-Q^2)^{1/2}}]-Q^2}{{(M^2-\frac{J^2}{M^2}-Q^2)^{1/2}}}\right),
\label{4.5}
\end{eqnarray}
\begin{eqnarray}
-\frac{\Omega_{\textrm H}}{T_{\textrm H}}=-\frac{2\pi J}{\hbar M}\left(\frac{1}{(M^2-\frac{J^2}{M^2}-Q^2)^{1/2}}\right),
\label{4.6}
\end{eqnarray}
\begin{eqnarray}
-\frac{\Phi_{\textrm H}}{T_{\textrm H}}=-\frac{2\pi Q[M+(M^2-\frac{J^2}{M^2}-Q^2)^{1/2}]}{\hbar(M^2-\frac{J^2}{M^2}-Q^2)^{1/2}}.
\label{4.7}
\end{eqnarray}

\subsection{Komar conserved quantities}
The Komar integral gives the conserved quantity corresponding to a Killing vector field. We take the following definition for the conserved quantities corresponding to the Killing fields $\partial_t$ and $\partial_{\phi}$ in Kerr-Newman spacetime, respectively, as {\footnote{ Our normalisation for $P_{\phi}$ is consistent with \cite{Katz}.}}
\begin{eqnarray}
E=\frac{1}{4\pi}\int_{\partial\Sigma}{d^{2}x\sqrt{\gamma^{(2)}}n_{\mu}\sigma_{\nu}\nabla^{\mu}K^{\nu}}
\label{komenergy}
\end{eqnarray}  
and
\begin{eqnarray}
P_{\phi}=-\frac{1}{4\pi}\int_{\partial\Sigma}{d^{2}x\sqrt{\gamma^{(2)}}n_{\mu}\sigma_{\nu}\nabla^{\mu}R^{\nu}}.
\label{komangm}
\end{eqnarray}   
The above two integrals are defined on the boundary ($\partial{\Sigma}$) of a spacelike hypersurface $\Sigma$ and $\gamma_{ij}$ is the induced metric on $\partial{\Sigma}$. Also, $n^{\mu}$ and $\sigma^{\nu}$ are unit normal vectors associated with $\Sigma$ and $\partial{\Sigma}$ respectively, whereas, $K^{\mu}$ and $R^{\nu}$ are timelike and rotational Killing vectors.

For the spherically symmetric spacetime (Schwarzschild and Reissner-Nordstrom) there is only one Killing vector ($\partial_t$) and correspondingly only one conserved quantity given by (\ref{komenergy}) $E_{\textrm {Sch}}= M$ and $E_{\textrm {RN}}= (M-\frac{Q^2}{r})$ respectively.

For the Kerr spacetime, in Boyer-Linquist coordinates ($t,~r,~\theta,~\phi$), $E_{\textrm K}= M$ and $P_{\phi}^{\textrm K}= 2J$.

For the Kerr-Newman black hole, in the evaluation of (\ref{komenergy}) and (\ref{komangm}), there will be extra contributions due to charge ($Q$) \cite{Cohen}. A closed form expression for ${P_{\phi}^{\textrm {KN}}}$ is not available. Calculating upto the leading  ${\cal O}(\frac{1}{r})$ we obtain $E_{\textrm {KN}}= (M-\frac{Q^2}{r})$ and $P_{\phi}^{\textrm {KN}}= 2(J-\frac{2Q^2a}{3r})$.\\\\

{\it {\bf {Acknowledgements:}}}
Authors wish to thank Debraj Roy for many useful discussions and technical help. One of the authors (S.K.M) thanks the  Council of Scientific and Industrial Research (C.S.I.R), Government of India, for financial support.


\begin{thebibliography}{99}
\bibitem{Bardeen}J.M. Bardeen, B. Carter, S.W. Hawking, {\it Commun. Math. Phys.} {\bf 31}, 161 (1973).
\bibitem{Beken}J.D.Bekenstein, {\it Lett. Nuovo Cimento} {\bf 4}, 737 (1972).\\
               J.D.Bekenstein, {\it Phys. Rev.} {\bf D 7}, 2333 (1973).\\
               J.D.Bekenstein, {\it Phys. Rev.} {\bf D 9}, 3292 (1974).
\bibitem{Hawk}S.W.Hawking,{\it Nature} {\bf 248}, 30 (1974).\\
              S.W.Hawking,{\it Commun. Math. Phys.} {\bf 43}, 199 (1975).\\
              S.W.Hawking, {\it Phys. Rev.} {\bf D 13}, 191 (1976).
\bibitem{Wilczek}M.K.Parikh and F.Wilczek, {\it Phys. Rev. Lett.} {\bf 85}, 5042 (2000) [arXiv:hep-th/9907001].\\
                 M.K.Parikh {\it Int. J. Mod. Phys.} {\bf D 13}, 2351 (2004) [arXiv:hep-th/0405160].
\bibitem{Cai}   M.Arzano, A.J.M.Medved and E.C.Vagenas, {\it JHEP} {\bf 0509}, 037 (2005) [arXiv:hep-th/0505266].\\
                A.J.M.Medved and E.C.Vagenas, {\it Mod. Phys. Lett.} {\bf A 20}, 2449 (2005) [arXiv:gr-qc/0504113].\\
                Qing-Quan Jiang, Shuang-Qing Wu and Xu Cai, {\it{Phys. Rev.}} {\bf{D 73}} 064003 (2006) [arXiv:hep-th/0512351].\\
                Yapeng Hu, Jingyi Zhang and Zheng Zhao, {\it{Mod. Phys. Lett.}} {\bf{A 21}} 2143 (2006) [arXiv:gr-qc/0611026].\\    
                Zhibo Xu and Bin Chen, {\it{Phys. Rev.}} {\bf{D 75}} 024041 (2007) [arXiv:hep-th/0612261].\\                               
                Cheng-Zhou Liu and Jian-Yang Zhu, [arXiv:gr-qc/0703055].\\
                B.D.Chowdhury, {\it{Pramana}} {\bf 70}, 593 (2008) [arXiv:hep-th/0605197].\\
                T.Pilling, {\it Phys. Lett.} {\bf B 660}, 402 (2008) [arXiv:0709.1624].\\
                Qing-Quan Jiang and Shuang-Qing Wu, {\it{Phys. Lett.}} {\bf{B 635}}, 151 (2006) [arXiv:hep-th/0511123].\\
                Ya-peng Hu, Jingyi Zhang and Zheng Zhao, {\it{Int.J.Mod.Phys.}} {\bf{D 16}}, 847 (2007) [arXiv:gr-qc/0611085].\\
                Jingyi Zhang, {\it Phys.Lett.} {\bf B}, 668, 353 (2008) [arXiv:0806.2441].\\
                K. Chiang, K. San-Min, P. Dan-Tao, F. Tsun, ``{\it Hawking radiation as tunneling and the first law of thermodynamics at apparent horizon in the FRW universe}'', arXiv:0812.3006 [hep-th].\\
                Ya-Peng Hu, Jing-Yi Zhang, Zheng Zhao, ``{\it A note on the Hawking radiation calculated by the quasi-classical tunneling method}'' arXiv:0901.2680 [gr-qc].
\bibitem{Paddy}K.Srinivasan and T.Padmanabhan, {\it Phys. Rev.} {\bf D 60}, 024007 (1999) [arxiv:gr-qc/9812028].\\ 
               S.Shankarnarayanan, K.Srinivasan and T.Padmanabhan, {\it Mod. Phys. Lett.} {\bf A 16}, 571 (2001) [arXiv:gr-qc/0007022].\\
               S.Shankarnarayanan, T.Padmanabhan, and K.Srinivasan, {\it Class. Quantum. Grav.} {\bf 19}, 2671 (2002)  [arXiv:gr-qc/0010042].\\
               S.Shankarnarayanan, {\it Phys. Rev.} {\bf D 67}, 084026 (2003) [arXiv:gr-qc/0301090].\\                 
               E.C.Vagenas, {\it Nuovo Cim.} {\bf B} 117, 899 (2002) [arXiv:hep-th/0111047].
\bibitem{Kern}  E.T.Akhmedov, V.Akhmedova and D.Singleton, {\it Phys. Lett.} {\bf B 642}, 124 (2006) [arXiv:hep-th/0608098].
                E.T.Akhmedov, V.Akhmedova, D.Singleton and T.Pilling, {\it Int.J.Mod.Phys.} {\bf A 22}, 1705 (2007) [arXiv:hep-th/0605137].\\
                R.D.Criscienzo and L.Vanzo, {\it Europhys.Lett.} {\bf 82}, 60001 (2008) [arXiv:0803.0435].\\
                P.Mitra, {\it Phys. Lett.} {\bf B 648}, 240 (2007) [arXiv:hep-th/0611265].\\
                H. M. Siahaan, Triyanta, ``{\it Hawking Radiation from a Vaidya Black Hole: A Semi-Classical Approach and Beyond}'', arXiv:0811.1132 [gr-qc].\\
                Tao Zhu, Ji-Rong Ren, ``{\it Quantum Corrections to Hawking Radiation for a FRW Universe}'', arXiv:0811.4074 [hep-th].\\
                R.Banerjee, B.R. Majhi, D. Roy,  ``{\it Corrections to Unruh effect in tunneling formalism and mapping with Hawking effect}'' arXiv:0901.0466 [hep-th].\\
                B.R. Majhi, S. Samanta, ``{\it Hawking Radiation due to Photon and Gravitino Tunneling}'', arXiv:0901.2258 [hep-th].
\bibitem{Majhiflux}R. Banerjee, B.R. Majhi, ``{\it Hawking black body spectrum from tunneling mechanism}'', {\it Phys. Lett.} {\bf B} ({\it In Press}), arXiv:0903.0250v1 [hep-th].
\bibitem{Fursaev}D.V.Fursaev, {\it Phys. Rev.} {\bf D 51}, R5352 (1995) [arXiv:hep-th/9412161].\\
                 R.B.Mann and S.N.Solodukhin, {\it Nucl. Phys.} {\bf B 523}, 293 (1998) [arXiv:hep-th/9709064].
\bibitem{Partha}R.K.Kaul and P.Majumdar, {\it Phys. Rev. Lett.} {\bf 84}, 5255 (2000) [arXiv:gr-qc/0002040].\\
                S Kloster, J Brannlund and A DeBenedicts, {\it Class. Quantum. Grav.}{\bf 25}, 065008 (2008) [arXiv:gr-qc/0702036].
\bibitem{Das} Saurya Das, Parthasarathi Majumdar, Rajat K. Bhaduri,{\it Class. Quant. Grav.} {\bf 19}, 2355  (2002) [arXiv:hep-th/0111001].\\
             S.S.More, {\it Class. Quantum Grav.} {\bf 22}, 4129 (2005) [gr-qc/0410071].\\
             S.Mukherjee and S.S.Pal, {\it JHEP} {\bf 0205}, 026 (2002) [arXiv:hep-th/0205164]. 
\bibitem{Carlip}S Carlip, {\it Class. Quantum Grav.} {\bf 17}, 4175 (2000) [arXiv: gr-qc/0005017].\\
                M.R. Setare, {\it Eur.Phys.J.C} {\bf 33}, 2004 [arXiv:hep-th/0309134] 
\bibitem{Hooft}G. t Hooft, {\it Int. J. Mod. Phys.} {\bf A} 11, 4623 (1996), [arXiv:gr-qc/9607022].\\
               S. Sarkar, S. Shankaranarayanan, L. Sriramkumar, {\it Phys.Rev.} {\bf D} 78, 024003 (2008), arXiv:0710.2013 [gr-qc]. 
\bibitem{R. Banerjee}R. Banerjee, B.R. Majhi, {\it Phys. Lett.} {\bf B 662}, (2008) 62 [arXiv:0801.0200].\\
                     R. Banerjee, B.R. Majhi and S.Samanta, {\it Phys. Rev.} {\bf D 77}, 124035 (2008) [arXiv:0801.3583].\\
                     R. Banerjee, B.R. Majhi and S.K. Modak, {\it Class. Quant. Grav.} {\bf 26}, 085010 (2009), [arXiv:0802.2176].                   
\bibitem{Majhi1}R.Banerjee, B.R.Majhi, {\it JHEP} {\bf 06}, 095, (2008) [arXiv:0805.2220].
\bibitem{Modak}S.K. Modak, {\it Phys. Lett.} {\bf B} {\bf 671}, 167 (2009) arXiv:0807.0959 [hep-th].
\bibitem{Majhitrace}R. Banerjee, B.R. Majhi, {\it Phys. Lett.} {\bf B 674}, 218 (2009), [arXiv:0808.3688].  
\bibitem{Kerner1}R.Kerner and R.B.Mann, {\it Phys. Rev.} {\bf D 73}, 104010 (2006) [arXiv:gr-qc/0603019].
\bibitem{Ang1} M.Angheben, M.Nadalini, L.Vanzo and S.Zerbini, {\it JHEP} {\bf 0505}, 014 (2005) [arXiv:hep-th/0503081].
\bibitem{Komar}A. Komar, {\it Phys. Rev.}, {\bf 113}, 3, (1959). 
\bibitem{Pilling}E.T. Akhmedhov, T. Pilling, D. Singleton, ``{\it Subtleties in the quasi-classical calculation of Hawking radiation}'' arXiv:0805.2653 [gr-qc]. 
\bibitem{Majhiconnect}R. Banerjee, B.R. Majhi, {\it Phys. Rev.} {\bf D} 79, 064024, (2009) arXiv:0812.0497 [hep-th]. 
\bibitem{fermion}R.Kerner and R.B.Mann, {\it Class. Quant. Grav.} {\bf 25}, 095014 (2008) arXiv:0710.0612 [hep-th].
\bibitem{fermoth}R. Li, Ji-R. Ren, {\it Phys.Lett.} {B 661} 370 (2008) [arXiv:0802.3954].\\ 
                 R.Di Criscienzo, L. Vanzo, {\it Europhys.Lett.} {\bf 82}, 60001 (2008) [arXiv:0803.0435].\\
                 R. Li, Ji-R. Ren, {\it Class.Quant.Grav.} {\bf 25} 125016, (2008) [arXiv:0803.1410].\\
                 R. Kerner, R.B. Mann, {\it Phys.Lett.} {\bf B665} 277, (2008) [arXiv:0803.2246].\\
                 De-Y. Chen, Qing-Q. Jiang, Shu-Z. Yang, Xiao-T. Zu, {\it Class.Quant.Grav.} {\bf 25} 205022 (2008) [arXiv:0803.3248].\\
                 De-Y. Chen, Qing-Q. Jiang, Xiao-T. Zu, {\it Phys.Lett.} {\bf B 665}, 106 (2008) [arXiv:0804.0131].\\ 
                 Qing-Q. Jiang, {\it Phys.Rev.} {\bf D78}, 044009 (2008) [arXiv:0807.1358].\\ 
                 Qing-Q. Jiang, {\it Phys.Lett.} {\bf B666}, 517 (2008).\\
                 R. Li, Ji-R. Ren, Dun-Fu Shi, {\it Phys.Lett.} {\bf B670}, 446 (2009) [arXiv:0812.4217].
\bibitem{Majhifermion} B. R. Majhi, {\it Phys. Rev.} {\bf D 79}, 044005 (2009) [arXiv:0809.1508].
\bibitem{Carrol}S.M.Carroll, ``An Introduction to General Relativity: Spacetime and Geometry'', {\it San Francisco, CA, USA: Addison Wesley}, (2004).             
\bibitem {Hawkzeta}S.W.Hawking, {\it Commun. Math. Phys.} {\bf 55}, 133 (1977).
\bibitem{Dewitt} Bryce S. DeWITT, {\it Phys. Rep.} {\bf 19}, 295 (1975).
\bibitem{Carrolpaper}S.M. Carrol, M.C. Johnson, and L. Randall, ``{\it Extremal limits and black hole entropy}'', arXiv:0901.0931 [hep-th].    
\bibitem{Katz}J Katz, {\it Class. Quant. Grav.} {\bf 2}, 423 (1985). 
\bibitem{Cohen}Jeffrey M. Cohen, F. de Felice, {\it J. Math. Phys.} {\bf 25}, 992 (1984).




\end{thebibliography}
\end{document}